\documentclass[]{aiaa-tc}

\usepackage{amsmath} 
\usepackage{bm} 
\usepackage{amssymb}
\usepackage{gensymb}
\usepackage{verbatim}
\usepackage{subfigure}
\usepackage{placeins}
\usepackage{overpic}

\usepackage[pdftex]{hyperref}
\hypersetup{
pdfauthor   = {},
pdftitle    = {},
pdfsubject  = {},
pdfkeywords = {},
pdfcreator  = {PDFLaTeX with hyperref package},
pdfproducer = {pdflatex},
colorlinks  = true,
linkcolor   = blue,
anchorcolor = red,
citecolor   = blue,
filecolor   = red,
urlcolor    = red
}

\title{Dynamic Mode Decomposition of High Reynolds Number Supersonic Jet Flows}

\author{
Sami Yamouni\thanks{Postdoctoral Reasearch Fellow,
Graduate Program on Computer Sciences and Electrical Engineering,
Departamento de Ci\^{e}ncia e Tecnologia Aeroespacial, DCTA/ITA;
E-mail: sami.yamouni@gmail.com.}\\
{\normalsize\itshape Instituto Tecnol\'{o}gico de Aeron\'{a}utica, 12228-900
S\~{a}o Jos\'{e} dos Campos, SP, Brazil}\\
\and
Carlos Junqueira-Junior\thanks{Postdoctoral Research Fellow,
Graduate Program on Computer Sciences and Electrical Engineering,
Departamento de Ci\^{e}ncia e Tecnologia Aeroespacial, DCTA/ITA;
E-mail: junior.hmg@gmail.com.}
\hspace*{0.05 cm} , \hspace*{0.05 cm}
Jo\~{a}o Luiz F. Azevedo
\thanks{Senior Research Engineer, Aerodynamics Division,
Departamento de Ci\^{e}ncia e Tecnologia Aeroespacial, DCTA/IAE/ALA;
E-mail: joaoluiz.azevedo@gmail.com. Fellow AIAA.}\\
{\normalsize\itshape Instituto de Aeron\'{a}utica e Espa\c{c}o, 12228-903
S\~{a}o Jos\'{e} dos Campos, SP, Brazil}
\and
William R. Wolf
\thanks{Assistant Professor, Faculty of Mechanical Engineering;
E-mail: wolf@fem.unicamp.br. Member AIAA.}\\
{\normalsize\itshape Universidade Estadual de Campinas, 13083-970
Campinas, SP, Brazil}
}

 \AIAApapernumber{2015-NUMBER}
 \AIAAconference{22nd AIAA Computational Fluid Dynamic Conference, June, 2015, Dallas, Texas}
 \AIAAcopyright{\AIAAcopyrightD{2015}}

\begin{document}

\maketitle


\begin{abstract}

\section*{Abstract}
Current design constraints have encouraged the studies of 
aeroacoustic fields around compressible jet flows. The 
present work addresses the numerical study of unsteady turbulent 
jet flows as a preparation for future aeroacoustic analyses of main 
engine rocket plumes. An in-house large eddy simulation tool is used 
in order to reproduce high fidelity results of compressible jet flows. 
The large eddy simulation formulation is written using a second order 
numerical scheme for a finite difference spatial discretization.
Numerical simulations of perfectly expanded jets are performed and the 
results are compared to the literature. Dynamic mode decompositions (DMD) 
of the jet flow, using large size three-dimensional snapshots, are performed.
Three variables are analyzed, namely, the velocity magnitude, the vorticity 
magnitude and the divergence of velocity. In particular, two frequencies 
are identified and they are linked to flow structures observed in experiments 
performed by other authors in the literature. The spatial shapes of the 
corresponding dynamic modes are also discussed.
\end{abstract}


  \section{Introduction}

One of the main design issues related to launch vehicles 
lies on noise emission originated by the complex 
interaction between the high-temperature/high-velocity 
exhaustion gases and the atmospheric air. These emissions, 
which have high noise levels, can damage the launching 
structure or even be reflected upon the vehicle 
structure itself and the equipment onboard at the top of 
the vehicles. Moreover, the resulting pressure fluctuations 
can damage the solid structure of different parts of the 
launcher or the onboard scientific equipment by 
vibrational acoustic stress. Therefore, it is strongly
recommended to consider the load resulted from acoustic 
sources over large launching vehicles during take-off 
and also during the transonic flight. Moreover, one cannot 
neglect the energy dissipation effect generated by the 
acoustic waves even if the vehicle is far from the ground. 
Theoretically, all chemical energy should be converted to 
kinetic energy. However, in reality, the noise generation 
consumes part of the chemical energy.

The acoustics design constraints have encouraged the 
studies of aeroacoustic fields around compressible jet 
flows. Instituto de Aeronautica e Espa\c{c}o (IAE), 
in Brazil, is interested in this flow configuration for 
rocket design applications. Unsteady property fields of 
the flow are necessary for the aerocoustic study. 
Therefore, the present work addresses the numerical 
study of unsteady turbulent compressible jet flows 
as a preparatory step for such aeroacoustic applications 
in the future. Large eddy simulations (LES) are used in 
order to reproduce high fidelity results of the unsteady 
compressible jet flows.
 

One issue with the results of such large data sets resulting from LES 
calculations is precisely the large amount of data that has to be handled.
Hence, with the objective of simplifying a complex flow into a low-dimensional representation
containing the dominant dynamic structures, the use of different techniques has been proposed.
Among these, Proper Orthogonal Decomposition (POD)\cite{lumley1970stochastic,sirovich1987turbulence,
berkooz1993proper} and Dynamic Mode Decomposition (DMD)\cite{schmid2010dynamic}
are the more commonly used techniques in the fluid dynamics community. 
POD selects the modes depending on their energy content. However, this criterion is not 
necessarily the most appropriate, since the energy is not always the key parameter in order 
to identify the flow structures\cite{schmid2010dynamic} of interest. In contrast with POD, 
the modes computed from the DMD approach define characteristic frequencies of the flow. 
Hence, DMD is the chosen method for the intended application of the present work. It has 
already been applied to various flow configurations, such as cavity 
flows\cite{schmid2010dynamic,seena2011dynamic}, shock wave--turbulent boundary layer 
interaction\cite{grilli2012analysis}, boundary layer flows\cite{sayadi2013dynamic,tang2012dynamic},
cylinder flows\cite{bagheri2013koopman,tissot2013dynamic}, combustion chamber 
flows\cite{jourdain2013application,abou2015cfd}, wake behind a flexible 
membrane\cite{schmid2010dynamic} or jet flows\cite{larusson2014investigation,schmid2011application,
schmid2011applications,schmid2012decomposition,stegeman2014dynamic,wan2015dynamic}.
Several variations of the DMD algorithm have also been proposed. One can cite the optimal 
mode decomposition\cite{wynn2013optimal}, the sparsity-promoting DMD\cite{jovanovic2014sparsity},
the extended DMD\cite{williams2014data,williams2015kernel}, or the streaming DMD\cite{hemati2014dynamic}\@.
Lately, an unbiased noise-robust method has been proposed by Hemati {\it et al.}\cite{hemati2015biasing} 
to overcome the adverse influence of measurement errors. This method can be combined with all the 
previously listed DMD algorithms.

Therefore, in this context, the main objective of the present work is to apply the DMD algorithm 
to the numerical data extracted from large eddy simulations of perfectly-expanded supersonic
jet flows at $M=1.4$\@. Due to the large dimension of this problem, the authors use the 
streaming version\cite{hemati2014dynamic} of the {\it total-least squares} DMD formulation proposed 
by Hemati {\it et al.}\cite{hemati2015biasing}
The DMD results are compared to numerical and experimental data available in the literature.



\section{Navier-Stokes Equations}

The numerical strategy used in the present study is based on the compressible 
Navier-Stokes equations formulated as
\begin{eqnarray}
\frac{\partial \rho}{\partial t} + \frac{\partial}{\partial x_{j}} \left( \rho u_{j} \right) = 0 
\, \mbox{,}\\
\frac{\partial}{\partial t} \left( \rho u_{i}\right) + \frac{\partial}{\partial x_{j}} 
\left( \rho u_{i} u_{j} \right)
+ \frac{\partial p}{\partial x_{i}} - \frac{\partial \tau_{ij}}{\partial x_{j}} = 0  \, \mbox{,} \\
\frac{\partial e}{\partial t} + \frac{\partial}{\partial x_{j}} \left[ \left( e + p \right)u_{j} 
- \tau_{ij}u_{i} + q_{j} \right] = 0 \, \mbox{,}
\label{eq:NS}
\end{eqnarray}
in which $t$ and $x_{i}$ are independent variables representing time and spatial coordinates of a 
Cartesian coordinate system $\textbf{x}$, respectively. The components of the velocity vector 
$\textbf{u}$ are written as $u_{i}$, and $i=1,2,3$. Density, pressure and total energy per mass unit 
are denoted by $\rho$, $p$ and $e$, respectively. The heat flux vector, $q_{j}$, is given by 
\begin{equation}
q_{j} = \kappa \frac{\partial T}{\partial x_{j}} \, \mbox{,}
\end{equation} 
where $T$ is the static temperature and $\kappa$ is the thermal conductivity coefficient, which 
can by expressed by
\begin{equation}
\kappa = \frac{\mu C_{p}}{Pr} \, \mbox{ .}
\end{equation}
The thermal conductivity coefficient is a function of the specific heat at constant pressure, 
$C_p$, of the Prandtl number, $Pr$, which is equal to $0.72$ for air, and of the dynamic 
viscosity coefficient, $\mu$. The latter can be calculated using Sutherland's law,
\begin{equation}
\mu \left( T \right) \, = \, \mu_{\infty} \left( \frac{T}{T_{\infty}} \right)^{\frac{3}{2}} 
\frac{T_{0}+S_{1}}{T+S_{1}} \mbox{ ,} \hspace*{1.0 cm}
\mbox{with} \: S_{1} = 110.4 \mbox{ K .}
\label{eq:sutherland}
\end{equation}
According to the Stokes hypothesis, the shear-stress tensor, $\tau_{ij}$, for a Newtonian fluid can be 
written as
\begin{equation}
\tau_{ij} = 2 \mu \left( S_{ij} - \frac{1}{3} \delta_{ij} S_{kk} \right) \, \mbox{,} 
\end{equation}
in which the components of rate-of-strain tensor, $S_{ij}$, are given by
\begin{equation}
S_{ij} = \frac{1}{2} \left( \frac{\partial u_{i}}{\partial x_{j}} 
       + \frac{\partial u_{j}}{\partial x_{i}} 
\right) \, \mbox{.}
\end{equation}

In order to close the system of equations the density, the static pressure and the static 
temperature are correlated by the equation of state given by
\begin{equation}
p = \rho R T \, \mbox{,}
\end{equation}
where $R$ is the gas constant, written as
\begin{equation}
R = C_{p} - C_{v} \, \mbox{,}
\end{equation}
and $C_{v}$ is the specif heat at constant volume. The total energy per mass unity is given by:
\begin{equation}
e = \frac{p}{\gamma - 1} + \frac{1}{2} \rho u_{i} u_{i} \, \mbox{,} 
\end{equation}
in which $\gamma$ is the ratio of specific heats, written as $\gamma = C_p/C_v$.

\section{Large Eddy Simulation Filtering}

The large eddy simulation is based on the principle of scale separation, which is addressed as a 
filtering procedure in a mathematical formalism. A modified version of the the System I filtering 
approach \cite{Vreman1995} is used in present work. The original formulation neglects the double 
correlation term and it is written as
\begin{equation}
\begin{array}{c}
\displaystyle \frac{\partial \overline{\rho} }{\partial t} + \frac{\partial}{\partial x_{j}} 
\left( \overline{\rho} \widetilde{  u_{j} } \right) = 0 \, \mbox{,}\\
\displaystyle \frac{\partial}{\partial t} \left( \overline{ \rho } \widetilde{ u_{i} } \right) 
+ \frac{\partial}{\partial x_{j}} 
\left( \overline{ \rho } \widetilde{ u_{i} } \widetilde{ u_{j} } \right)
+ \frac{\partial \overline{p}}{\partial x_{i}} 
- \frac{\partial \check{\tau}_{ij}}{\partial x_{j}} = 
-\frac{\partial}{\partial x_{j}} \left[ \sigma_{ij} - \left( \overline{\tau_{ij}} 
- \check{\tau}_{ij}  \right) \right] \, \mbox{,} \\ 
\displaystyle \frac{\partial \check{e}}{\partial t} 
+ \frac{\partial}{\partial x_{j}}
\left[ \left( \check{e} + \overline{p} \right)\widetilde{u_{j}} \right]
- \frac{\partial \check{\tau}_{ij} \widetilde{u_{i}} }{\partial x_{j}} 
+ \frac{\partial \check{q}_{j}}{\partial x_{j}} =
-B_{1}-B_{2}-B_{3}+B_{4}+B_{5}+B_{6}-B_{7} \, \mbox{.}
\end{array}
\label{eq:full_system_I}
\end{equation}
The $\left( \tilde{\cdot} \right)$ notation is used to represent a Frave averaged property. The 
SGS stress tensor components are written as $\sigma_{ij}$. The filtering procedure originates two 
new terms, $\check{\tau}_{ij}$ and $\check{q}_{j}$. These new terms are given by
\begin{equation}
	\check{\tau}_{ij} = 2 \mu \left( \tilde{S}_{ij} 
	                  - \frac{1}{3} \delta_{ij} \tilde{S}_{kk} \right) 
	\, \mbox{,} 
\end{equation}
where
\begin{equation}
	\tilde{S}_{ij} = \frac{1}{2} \left( \frac{\partial \tilde{u}_{i}}{\partial x_{j}} 
	+ \frac{\partial \tilde{u}_{j}}{\partial x_{i}} 
\right) \, \mbox{,}
\end{equation}
and
\begin{equation}
	\check{q}_{j} = \check{\kappa} \frac{\partial \tilde{T}}{\partial x_{j}} \, \mbox{,}
\end{equation} 
in which
\begin{equation}
	\check{\kappa} = \kappa \left(\tilde{T}\right)
	               = \frac{\tilde{\mu}\left(\tilde{T}\right) C_{p}}{Pr} \, \mbox{.}
\end{equation}
The SGS terms of the energy equation, $B_{i}$, are given by
\begin{eqnarray}
B_{1} = \frac{1}{(\gamma - 1)} \frac{\partial}{\partial x_{j}} \left( \overline{p u_{j}} 
- \overline{p} \widetilde{u_{j}} \right) 
= \frac{\partial C_{v} Q_{j}}{\partial x_{j}} \, \mbox{,} \\
B_{2} = \overline{p \frac{\partial u_{k}}{\partial x_{k}} } 
- \overline{p} \frac{\partial \widetilde{u_{k}}}{\partial x_{k}} 
= \Pi_{dil} \, \mbox{,}\\
B_{3} = \frac{\partial}{\partial x_{j}} \left( \sigma_{kj} \widetilde{u_{k}} \right) \, \mbox{,} \\
B_{4} = \sigma_{kj} \frac{\partial}{\partial x_{j}} \widetilde{u_{k}} \, \mbox{,} \\
B_{5} = \overline{\tau_{kj} \frac{\partial}{\partial x_{j}} u_{k}} - \overline{\tau_{ij}} 
\frac{\partial}{\partial x_{j}}\widetilde{u_{k}} 
= \epsilon \, \mbox{,} \\
B_{6} = \frac{\partial}{\partial x_{j}} \left( \overline{\tau_{ij}}\widetilde{u_{i}} 
- \check{\tau}_{ij}\widetilde{u_{i}} \right) 
= \frac{\partial \mathcal{D}}{\partial x_{j}} \, \mbox{,} \\
B_{7} = \frac{\partial}{\partial x_{j}} \left( \overline{q_{j}} - \check{q}_{j} \right) \, \mbox{.}
\end{eqnarray}
The work of Vreman {\it et al.} \cite{Vreman95} and Vreman \cite{Vreman1995} classify the 
influence of each term of System I and System II formulations on a 2-D temporal shear layer flow.
The classification, including large, medium, small and negligible effects, is based on 
the $L_{2}$ norm of different terms of the filtered equations. One order of magnitude separates the 
norm of each class of terms. Garnier {\it et al.} \cite{Garnier09} compile the analogy as presented 
in Tab.\ \ref{tab:class_syst_I}.
\begin{table}[htbp!]
\begin{center}
\caption{Classification of System I terms}
\label{tab:class_syst_I}
\begin{tabular}{cc}
\hline\hline
& \\
Large & convective $\overline{NS}$ \\
Medium & diffusive $\overline{NS}$, $B_{1}$, $B_{2}$ and $B_{3}$ \\
Small & $B_{4}$ and $B_{5}$\\
Negligible & $\frac{\partial}{\partial x_{j}} \left( \overline{\tau_{ij}} - \check{\tau_{ij}} \right)$,
$B_{6}$ and $B_{7}$ \\
& \\
\hline\hline
\end{tabular}
\end{center}
\end{table}

In practice, the authors of the System I analogy neglect the non-linear terms occuring in the viscous terms
and in the heat fluxes \cite{Garnier09}. Moreover, some of the terms from the original 
System I set of equations, Eq.\ \eqref{eq:full_system_I}, such as $B_{4}$ and $B_{5}$, cannot be 
written in conservative form. Only the terms with large and medium influence are considered in 
the present work. The SGS stress tensor components are written using the SGS viscosity 
\cite{Sagaut05},
\begin{equation}
    \sigma_{ij} = - 2 \mu_{sgs} \left( \check{S}_{ij} - \frac{1}{3} \check{S}_{kk} \right)
    + \frac{1}{3} \delta_{ij} \sigma_{kk}
    \, \mbox{.}
    \label{eq:sgs_visc}
\end{equation}
The most important terms of the filtered energy equation are modeled based on the work of Eidson
\cite{eidson85} and Vreman \cite{Vreman1995}. They are given by
\begin{equation}
	B_{1}+B_{2} = - \frac{\partial}{\partial x_{j}} 
	\left( \kappa_{sgs} \frac{\partial \tilde{T}}{\partial x_{j}} \right)
	\, \mbox{,}
	\label{eq:b1_b2_model}
\end{equation}
where
\begin{equation}
	\kappa_{sgs} = \frac{\mu_{sgs} C_{p}}{ {Pr}_{sgs} } \, \mbox{.}
	\label{eq:kappa_sgs}
\end{equation}
Using Eqs.\ \eqref{eq:sgs_visc}, \eqref{eq:b1_b2_model}, and \eqref{eq:kappa_sgs}, 
one can write a simplified version of the System I formulation as
\begin{equation}
\begin{array}{c}
\displaystyle \frac{\partial \overline{\rho} }{\partial t} + \frac{\partial}{\partial x_{j}} 
\left( \overline{\rho} \widetilde{  u_{j} } \right) = 0 \, \mbox{,}\\
\displaystyle \frac{\partial}{\partial t} \left( \overline{ \rho } \widetilde{ u_{i} } \right) 
+ \frac{\partial}{\partial x_{j}} 
\left( \overline{ \rho } \widetilde{ u_{i} } \widetilde{ u_{j} } \right)
+ \frac{\partial \overline{p}}{\partial x_{i}} 
- \frac{\partial {\tau}^{mod}_{ij}}{\partial x_{j}}  
+ \frac{1}{3} \frac{\partial}{\partial x_{j}}\left( {\delta}_{ij} \sigma_{ii}\right)
= 0 \, \mbox{,} \\ 
\displaystyle \frac{\partial \check{e}}{\partial t} 
+ \frac{\partial}{\partial x_{j}} 
\left[ \left( \check{e} + \overline{p} \right)\widetilde{u_{j}} \right]
- \frac{\partial}{\partial x_{j}}\left({\tau}^{mod}_{ij} \widetilde{u_{i}} \right)
+ \frac{1}{3} \frac{\partial}{\partial x_{j}}
\left[ \left( \delta_{ij}{\sigma}_{ii} \right) \widetilde{u_{i}} \right]
+ \frac{\partial {q}^{mod}_{j}}{\partial x_{j}} = 0 \, \mbox{,}
\end{array}
\label{eq:modified_system_I}
\end{equation}
where, ${\tau}^{mod}_{ij}$ and ${q}^{mod}_{j}$, include the viscous and the subgrid terms.
They can be written as
\begin{equation}
	{\tau}^{mod}_{ij} = 2 \left(\mu+{\mu}_{sgs}\right) 
	\left( S_{ij} - \frac{1}{3} \delta_{ij} S_{kk} \right) \,
	\label{eq:tau_mod}
\end{equation}
and
\begin{equation}
	{q}^{mod}_{j} = \left(\kappa+{\kappa}_{sgs}\right) \frac{\partial T}{\partial x_{j}} 
	\, \mbox{.}
	\label{eq:q_mod}
\end{equation}

Previous work have shown that the subgrid scale terms are too small when compared to 
the truncation error of the second order numerical scheme used in the current research
\cite{Junior15, Junior16, Junior16PhD}. Therefore, an implicit LES is performed in which 
all subgrid scales terms, $[\cdot]_{sgs}$, introduced in Eqs.\ \eqref{eq:tau_mod} and 
\eqref{eq:q_mod} are neglected.

\section{Transformation of Coordinates}

The formulation used in the current work is written in the a general curviliar coordinate 
system in order to facilitate the implementation and add more generality for the CFD tool. 
The modified System I set of equations, Eq.\ \eqref{eq:modified_system_I} can be written 
in a strong conservative form for a 3-D Cartesian coordinate system as
\begin{equation}
	\frac{\partial \overline{Q}}{\partial t} + \frac{\partial \overline{E}}{\partial x} +
	\frac{\partial \overline{F}}{\partial y} + \frac{\partial \overline{G}}{\partial z} = 0 \, \mbox{,}
	\label{eq:vec-RANS}
\end{equation}
where $\overline{Q}$ stands for the filtered conservative properties vector given by
\begin{equation}
	\overline{Q} = \left[ \overline{\rho} \quad \overline{\rho}\tilde{u} \quad 
	\overline{\rho}\tilde{v} \quad \overline{\rho}\tilde{w} \quad \check{e} \right]^{T} \quad \mbox{.}
\end{equation}
The flux vectors which represent both the inviscid and viscous fluxes, $\overline{E}$, $\overline{F}$ 
and $\overline{G}$ are written as
\begin{equation}
	\overline{E} = \left\{
	\begin{array}{c}
		\overline{\rho}\tilde{u} \\
		\overline{\rho}\tilde{u}^{2} + \overline{p} - {\tau}^{mod}_{xx} + \frac{1}{3}\sigma_{xx}\\
		\overline{\rho}\tilde{u}\tilde{v} - {\tau}^{mod}_{xy} \\
		\overline{\rho}\tilde{u}\tilde{w} - {\tau}^{mod}_{xz} \\
		\left( \check{e} + \overline{p} - {\tau}^{mod}_{xx} + \frac{1}{3}\sigma_{xx}\right) \tilde{u} -
		{\tau}^{mod}_{xy}\tilde{v} - {\tau}^{mod}_{xz}\tilde{w} + q^{mod}_{x}
	\end{array} \right\} \, \mbox{,}
\end{equation}
\begin{equation}
	\overline{F} = \left\{
	\begin{array}{c}
		\overline{\rho}\tilde{v} \\
		\overline{\rho}\tilde{u}\tilde{v} - {\tau}^{mod}_{xy} \\
		\overline{\rho}\tilde{v}^{2} + \overline{p} - {\tau}^{mod}_{yy} + \frac{1}{3}\sigma_{yy} \\
		\overline{\rho}\tilde{v}\tilde{w} - {\tau}^{mod}_{yz} \\
		\left( \check{e} + \overline{p} - {\tau}^{mod}_{yy} + \frac{1}{3}\sigma_{yy}\right) \tilde{v} -
		{\tau}^{mod}_{xy}\tilde{u} - {\tau}^{mod}_{yz}\tilde{w} + q^{mod}_{y}
	\end{array} \right\} \, \mbox{,}
\end{equation}
\begin{equation}
	\overline{G} = \left\{
	\begin{array}{c}
		\overline{\rho}\tilde{w} \\
		\overline{\rho}\tilde{u}\tilde{w} - {\tau}^{mod}_{xz} \\
		\overline{\rho}\tilde{v}\tilde{w} - {\tau}^{mod}_{yz} \\
		\overline{\rho}\tilde{w}^{2} + \overline{p} - {\tau}^{mod}_{zz} + \frac{1}{3}\sigma_{zz}\\
		\left( \check{e} + \overline{p} - {\tau}^{mod}_{zz} + \frac{1}{3}\sigma_{zz}\right) \tilde{w} -
		{\tau}^{mod}_{xz}\tilde{u} - {\tau}^{mod}_{yz}\tilde{v} + q^{mod}_{z}
	\end{array} \right\} \, \mbox{,} 
\end{equation}
in which, $u$, $v$ and $w$ are the velocity components in the Cartesian coordinates, 
$x$, $y$ and $z$ respectively.

In the present work the chosen general coordinate transformation is given by
\begin{eqnarray}
	\mathcal{T} & = & t \, \mbox{,} \nonumber\\
	\xi & = & \xi \left(x,y,z,t \right)  \, \mbox{,} \nonumber\\
	\eta & = & \eta \left(x,y,z,t \right)  \, \mbox{,} \\
	\zeta & = & \zeta \left(x,y,z,t \right)\ \, \mbox{.} \nonumber
\end{eqnarray}
Throughout the present work, $\xi$ is the axial jet flow direction, $\eta$ is the 
radial direction and $\zeta$ is the azimuthal direction.
The derivatives in the general curvilinear coordinate system are calculated as a function of the 
derivatives the Cartesian coordinate system by the chain rule. Therefore, one can write
\begin{equation}
	\left\{\begin{array}{c}
		\frac{\partial}{\partial \mathcal{T}} \\
		\frac{\partial}{\partial \xi} \\
		\frac{\partial}{\partial \eta} \\
		\frac{\partial}{\partial \zeta}
	\end{array} \right\} =
	\left[\begin{array}{cccc}
		1 & x_{\mathcal{T}} & y_{\mathcal{T}} & z_{\mathcal{T}} \\
		0 & x_{\xi} & y_{\xi} & z_{\xi} \\
		0 & x_{\eta} & y_{\eta} & z_{\eta} \\
		0 & x_{\zeta} & y_{\zeta} & z_{\zeta} \\
	\end{array}\right]
	\left\{\begin{array}{c}
		\frac{\partial}{\partial t} \\
		\frac{\partial}{\partial x} \\
		\frac{\partial}{\partial y} \\
		\frac{\partial}{\partial z}
	\end{array} \right\} \, \mbox{.}
	\label{eq:transf-matrix}
\end{equation}
The Jacobian of the transformation, $J$, is calculated as the inverse of the determinant of the matrix
in the chain rule presented in Eq.\ \eqref{eq:transf-matrix}. Therefore, for the 3-D coordinate 
transformation, the Jacobian can be written as
\begin{equation}
	J = \left( x_{\xi} y_{\eta} z_{\zeta} + x_{\eta}y_{\zeta}z_{\xi} +
	           x_{\zeta} y_{\xi} z_{\eta} - x_{\xi}y_{\zeta}z_{\eta} -
			   x_{\eta} y_{\xi} z_{\zeta} - x_{\zeta}y_{\eta}z_{\xi} 
	    \right)^{-1} \, \mbox{.}
\end{equation}
The metric terms are given by
\begin{eqnarray}
	\xi_{x} = J \left( y_{\eta}z_{\zeta} - y_{\zeta}z_{\eta} \right) \, \mbox{,} & 
	\xi_{y} = J \left( z_{\eta}x_{\zeta} - z_{\zeta}x_{\eta} \right) \, \mbox{,} & 
	\xi_{z} = J \left( x_{\eta}y_{\zeta} - x_{\zeta}y_{\eta} \right) \, \mbox{,} 
	\nonumber \\
	\eta_{x} = J \left( y_{\eta}z_{\xi} - y_{\xi}z_{\eta} \right) \, \mbox{,} & 
	\eta_{y} = J \left( z_{\eta}x_{\xi} - z_{\xi}x_{\eta} \right) \, \mbox{,} & 
	\eta_{z} = J \left( x_{\eta}y_{\xi} - x_{\xi}y_{\eta} \right) \, \mbox{,} \\
	\zeta_{x} = J \left( y_{\xi}z_{\eta} - y_{\eta}z_{\xi} \right) \, \mbox{,} & 
	\zeta_{y} = J \left( z_{\xi}x_{\eta} - z_{\eta}x_{\xi} \right) \, \mbox{,} & 
	\zeta_{z} = J \left( x_{\xi}y_{\eta} - x_{\eta}y_{\xi} \right) \, \mbox{,} 
	\nonumber \\
	\xi_{t} = -x_{\mathcal{T}}\xi_{x} - y_{\mathcal{T}}\xi_{y} - z_{\mathcal{T}}\xi_{z} 
	\, \mbox{,} & 
	\eta_{t} = -x_{\mathcal{T}}\eta_{x} - y_{\mathcal{T}}\eta_{y} - z_{\mathcal{T}}\eta_{z} 
	\, \mbox{,} & 
	\zeta_{t} = -x_{\mathcal{T}}\zeta_{x} - y_{\mathcal{T}}\zeta_{y} - z_{\mathcal{T}}\zeta_{z} 
	\, \mbox{.} 
	\nonumber
\end{eqnarray}

One can rewrite Eq.\ \eqref{eq:vec-RANS}, in a conservative form, for the general curvilinear 
coordinate system as
\begin{equation}
	\frac{\partial \hat{Q}}{\partial \mathcal{T}} + \frac{\partial \hat{E}}{\partial \xi} +
	\frac{\partial \hat{F}}{\partial \eta} + \frac{\partial \hat{G}}{\partial \zeta} = 0 \, \mbox{,}
	\label{eq:vec-hat-RANS}
\end{equation}
where 
\begin{equation}
	\hat{Q} = J^{-1}\overline{Q} = J^{-1} \left[ \overline{\rho} \quad \overline{\rho}\tilde{u} \quad 
	\overline{\rho}\tilde{v} \quad \overline{\rho}\tilde{w} \quad \check{e} \right]^{T} \quad \mbox{,}
	\label{eq:hat_Q_vec}
\end{equation}
and the new flux vectors are given by
\begin{eqnarray}
	\hat{E} = J^{-1} \left( 
	\xi_{t}\overline{Q} + \xi_{x}\overline{E} + \xi_{y}\overline{F} + \xi_{z}\overline{G} 
	\right) \, \mbox{,} \nonumber \\
	\hat{F} = J^{-1} \left( 
	\eta_{t}\overline{Q} + \eta_{x}\overline{E} + \eta_{y}\overline{F} + \eta_{z}\overline{G} 
	\right) \, \mbox{,} \\  
	\hat{G} = J^{-1} \left( 
	\zeta_{t}\overline{Q} + \zeta_{x}\overline{E} + \zeta_{y}\overline{F} + \zeta_{z}\overline{G} 
	\right) \, \mbox{.} \nonumber
\end{eqnarray}

Finally, the flux vectors are split in inviscid and viscous part in order to simplify the 
implementation. Therefore, Eq.\ \eqref{eq:vec-hat-RANS} can be rewritten as
\begin{equation}
	\frac{\partial \hat{Q}}{\partial \mathcal{T}} + \frac{\partial \hat{E}_{e}}{\partial \xi} +
	\frac{\partial \hat{F}_{e}}{\partial \eta} + \frac{\partial \hat{G}_{e}}{\partial \zeta} = 
	\frac{\partial \hat{E}_{v}}{\partial \xi} + \frac{\partial \hat{F}_{v}}{\partial \eta} + 
	\frac{\partial \hat{G}_{v}}{\partial \zeta} \, \mbox{,}
	\label{eq:vec-hat-split-RANS}
\end{equation}
where the inviscid flux vectors, $\hat{E}_{e}$, $\hat{F}_{e}$ and $\hat{G}_{e}$, are given by
\begin{equation}
	\hat{E}_{e} = J^{-1} \left\{\begin{array}{c}
		\overline{\rho} U \\
		\overline{\rho}\tilde{u} U + \overline{p} \xi_{x} \\
		\overline{\rho}\tilde{v} U + \overline{p} \xi_{y} \\
		\overline{\rho}\tilde{w} U + \overline{p} \xi_{z} \\
		\left( \check{e} + \overline{p} \right) U - \overline{p} \xi_{t}
	\end{array}\right\} \, \mbox{,}
	\label{eq:hat-flux-E}
\end{equation}
\begin{equation}
	\hat{F}_{e} = J^{-1} \left\{\begin{array}{c}
		\overline{\rho} V \\
		\overline{\rho}\tilde{u} V + \overline{p} \eta_{x} \\
		\overline{\rho}\tilde{v} V + \overline{p} \eta_{y} \\
		\overline{\rho}\tilde{w} V + \overline{p} \eta_{z} \\
		\left( \check{e} + \overline{p} \right) V - \overline{p} \eta_{t}
	\end{array}\right\} \, \mbox{,}
	\label{eq:hat-flux-F}
\end{equation}
\begin{equation}
	\hat{G}_{e} = J^{-1} \left\{\begin{array}{c}
		\overline{\rho} W \\
		\overline{\rho}\tilde{u} W + \overline{p} \zeta_{x} \\
		\overline{\rho}\tilde{v} W + \overline{p} \zeta_{y} \\
		\overline{\rho}\tilde{w} W + \overline{p} \zeta_{z} \\
		\left( \check{e} + \overline{p} \right) W - \overline{p} \zeta_{t}
	\end{array}\right\} \, \mbox{,}
	\label{eq:hat-flux-G}
\end{equation}
in which the contravariant velocity components, $U$, $V$ and $W$, are calculated as
\begin{eqnarray}
  U = \xi_{t} + \xi_{x}\overline{u} + \xi_{y}\overline{v} + \xi_{z}\overline{w} \, \mbox{,} \nonumber \\
  V = \eta_{t} + \eta_{x}\overline{u} + \eta_{y}\overline{v} + \eta_{z}\overline{w} \, \mbox{,} \\
  W = \zeta_{t} + \zeta_{x}\overline{u} + \zeta_{y}\overline{v} + \zeta_{z}\overline{w} \, \mbox{.} \nonumber
  \label{eq:vel_contrv}
\end{eqnarray}
The viscous flux vectors, $\hat{E}_{v}$, $\hat{F}_{v}$ and $\hat{G}_{v}$, are written as
\begin{equation}
	\hat{E}_{v} = J^{-1} \left\{\begin{array}{c}
		0 \\
		\xi_{x}\left({\tau}^{mod}_{xx}-\frac{1}{3}\sigma_{xx}\right) 
		+  \xi_{y}{\tau}^{mod}_{xy} + \xi_{z}{\tau}^{mod}_{xz} \\
		\xi_{x}{\tau}^{mod}_{xy} + \xi_{y}\left({\tau}^{mod}_{yy}-\frac{1}{3}\sigma_{yy}\right) 
		+ \xi_{z}{\tau}^{mod}_{yz} \\
		\xi_{x}{\tau}^{mod}_{xz} +  \xi_{y}{\tau}^{mod}_{yz} 
		+ \xi_{z}\left({\tau}^{mod}_{zz}-\frac{1}{3}\sigma_{zz}\right) \\
		\xi_{x}{\beta}_{x} +  \xi_{y}{\beta}_{y} + \xi_{z}{\beta}_{z} 
	\end{array}\right\} \, \mbox{,}
	\label{eq:hat-flux-Ev}
\end{equation}
\begin{equation}
	\hat{F}_{v} = J^{-1} \left\{\begin{array}{c}
		0 \\
		\eta_{x}\left({\tau}^{mod}_{xx}-\frac{1}{3}\sigma_{xx}\right) 
		+  \eta_{y}{\tau}^{mod}_{xy} + \eta_{z}{\tau}^{mod}_{xz} \\
		\eta_{x}{\tau}^{mod}_{xy} + \eta_{y}\left({\tau}^{mod}_{yy}-\frac{1}{3}\sigma_{yy}\right) 
		+ \eta_{z}{\tau}^{mod}_{yz} \\
		\eta_{x}{\tau}^{mod}_{xz} + \eta_{y}{\tau}^{mod}_{yz} 
		+ \eta_{z}\left({\tau}^{mod}_{zz}-\frac{1}{3}\sigma_{zz}\right) \\
		\eta_{x}{\beta}_{x} +  \eta_{y}{\beta}_{y} + \eta_{z}{\beta}_{z} 
	\end{array}\right\} \, \mbox{,}
	\label{eq:hat-flux-Fv}
\end{equation}
\begin{equation}
	\hat{G}_{v} = J^{-1} \left\{\begin{array}{c}
		0 \\
		\zeta_{x}\left({\tau}^{mod}_{xx}-\frac{1}{3}\sigma_{xx}\right) 
		+  \zeta_{y}{\tau}^{mod}_{xy} + \zeta_{z}{\tau}^{mod}_{xz} \\
		\zeta_{x}{\tau}^{mod}_{xy} +  \zeta_{y}\left({\tau}^{mod}_{yy}-\frac{1}{3}\sigma_{yy}\right) 
		+ \zeta_{z}{\tau}^{mod}_{yz} \\
		\zeta_{x}{\tau}^{mod}_{xz} +  \zeta_{y}{\tau}^{mod}_{yz} 
		+ \zeta_{z}\left({\tau}^{mod}_{zz}-\frac{1}{3}\sigma_{zz}\right) \\
		\zeta_{x}{\beta}_{x} +  \zeta_{y}{\beta}_{y} + \zeta_{z}{\beta}_{z} 
	\end{array}\right\} \, \mbox{,}
	\label{eq:hat-flux-Gv}
\end{equation}
where $\beta_{x}$, $\beta_{y}$ and $\beta_{z}$ are defined as
\begin{eqnarray}
	\beta_{x} = \left({\tau}^{mod}_{xx}-\frac{1}{3}\sigma_{xx}\right)\tilde{u} 
	+ {\tau}^{mod}_{xy}\tilde{v} + {\tau}^{mod}_{xz}\tilde{w} 
	- {q}^{mod}_{x} \, \mbox{,} \nonumber \\
	\beta_{y} = {\tau}^{mod}_{xy}\tilde{u} 
	+ \left({\tau}^{mod}_{yy}-\frac{1}{3}\sigma_{yy}\right)\tilde{v} +
	{\tau}^{mod}_{yz}\tilde{w} - {q}^{mod}_{y} \, \mbox{,} \\
	\beta_{z} = {\tau}^{mod}_{xz}\tilde{u} + {\tau}^{mod}_{yz}\tilde{v} +
	\left({\tau}^{mod}_{zz}-\frac{1}{3}\sigma_{zz}\right)\tilde{w} 
	- {q}^{mod}_{z} \mbox{.} \nonumber
\end{eqnarray}

\section{Dimensionless LES Formulation}

A convenient nondimensionalisation is necessary in order to achieve a consistent 
implementation of the governing equations of motion. Dimensionless formulation 
yelds to a more general numerical tool. There is no need to change the formulation 
for each configuration intended to be simulated. Moreover, dimensionless formulation 
scales all the necessary properties to the same order of magnitude which is a 
computational advantage \cite{BIGA02}. Dimensionless variables are presented in the 
present section in order perform the nondimensionalisation of Eq.\ 
\eqref{eq:vec-hat-split-RANS}.

The dimensionless time, $\underline{\mathcal{T}}$, is written as function of the 
freestream speed of sound and of a reference lenght, $D$,
\begin{equation}
	\underline{\mathcal{T}} = \mathcal{T} \frac{a_{\infty}}{l} \, \mbox{.}
	\label{eq:non-dim-time}
\end{equation}
In the current work, $D$ represents the jet entrance diameter. This reference 
lengh is aldo applied to write the dimensionless length, 
\begin{equation}
	\underline{l} = \frac{l}{D} \, \mbox{.}
	\label{eq:non-dim-lengh}
\end{equation}
The dimensionless velocity components are obtained using the freestream speed of 
sound
\begin{eqnarray}
	\underline{vel} = \frac{v}{a_{\infty}} & vel = u, v, w \, \mbox{.}
	\label{eq:non-dim-vel}
\end{eqnarray}
Dimensionless pressure and energy are calculated as
\begin{equation}
	\underline{p} = \frac{p}{\rho_{\infty}a_{\infty}^{2}} \, \mbox{,}
	\label{eq:non-dim-press}
\end{equation}
\begin{equation}
	\underline{e} = \frac{e}{\rho_{\infty}a_{\infty}^{2}} \, \mbox{.}
	\label{eq:non-dim-energy}
\end{equation}
Dimensionless density, $\underline{\rho}$, temperature, $\underline{T}$ and 
viscosity, $\underline{\mu}$, are calculated using freestream properties
\begin{equation}
	\underline{\rho} = \frac{\rho}{\rho_{\infty}} \, \mbox{.}
	\label{eq:non-dim-rho}
\end{equation}

One can use the dimensionless properties described above in order to write the 
dimensionless form of the LES equations as
\begin{equation}
	\frac{\partial \underline{\hat{Q}}}{\partial \underline{\mathcal{T}}} 
	+ \frac{\partial \underline{\hat{E}}_{e}}{\partial \underline{\xi}} +
	\frac{\partial \underline{\hat{F}}_{e}}{\partial \underline{\eta}} 
	+ \frac{\partial \underline{\hat{G}}_{e}}{\partial \underline{\zeta}} =
	\frac{M_{j}}{Re} \left( \frac{\partial \underline{\hat{E}}_{v}}{\partial \underline{\xi}} 
	+ \frac{\partial \underline{\hat{F}}_{v}}{\partial \underline{\eta}} 
	+ \frac{\partial \underline{\hat{G}}_{v}}{\partial \underline{\zeta}} \right)	\, \mbox{,}
	\label{eq:vec-underline-split-RANS}
\end{equation}
where the underlined terms are calculated using non dimensional properties. The jet Mach and Reynolds 
numbers are based on the mean jet inlet velocity, $U_{j}$, the freestream speed of sound, $a_{\infty}$, 
density, $\rho_{\infty}$, viscosity, $\mu_{\infty}$ and the reference length, $D$,
\begin{equation}
	M_{j} = \frac{U_{j}}{a_{\infty}} \mbox{ ,} \hspace*{2.0 cm}
	Re = \frac{\rho_{\infty}U_{j} D}{\mu_{\infty}} \, \mbox{.}
\end{equation}
%


  \section{Numerical Formulation}


The governing equations previously described are discretized in a 
structured finite difference context for general curvilinear 
coordinate system \cite{BIGA02}. The numerical flux is calculated 
through a central difference scheme with the explicit addition 
of the anisotropic artificial dissipation of Turkel and Vatsa
\cite{Turkel_Vatsa_1994}. The time integration is performed by an 
explicit, 2nd-order, 5-stage Runge-Kutta scheme 
\cite{jameson_mavriplis_86, Jameson81}.  Conserved properties
and artificial dissipation terms are properly treated near boundaries in order
to assure the physical correctness of the numerical formulation. 
 
\subsection{Spatial Discretization}

For the sake of simplicity, the formulation discussed in the present section
is no longer written using bars, underbars, etc. However, the reader should notice that the 
equations are dimensionless and filtered. The LES equations, presented in Eq.\ 
\eqref{eq:vec-underline-split-RANS}, are discretized in space in a finite 
difference fashion and, then, rewritten as
\begin{equation}
	\left(\frac{\partial Q}{\partial \mathcal{T}}\right)_{i,j,k} \  
	= \  - RHS_{i,j,k} \, \mbox{,}	
	\label{eq:spatial_discret}
\end{equation}
where $RHS$ is the right hand side of the equation and it is written as function of 
the numerical flux vectors at the interfaces between grid points,
\begin{eqnarray}
	{RHS}_{i,j,k} & = & 
	\frac{1}{\Delta \xi} \left( 
	{\mathbf{E}_{e}}_{(i+\frac{1}{2},j,k)} - {\mathbf{E}_{e}}_{(i-\frac{1}{2},j,k)} - 
	{\mathbf{E}_{v}}_{(i+\frac{1}{2},j,k)} + {\mathbf{E}_{v}}_{(i-\frac{1}{2},j,k)} 
	\right) \nonumber \\
	& & \frac{1}{\Delta \eta} \left( 
	{\mathbf{F}_{e}}_{(i,j+\frac{1}{2},k)} - {\mathbf{F}_{e}}_{(i,j-\frac{1}{2},k)} - 
	{\mathbf{F}_{v}}_{(i,j+\frac{1}{2},k)} + {\mathbf{F}_{v}}_{(i,j-\frac{1}{2},k)} 
	\right) \\
	& & \frac{1}{\Delta \zeta} \left( 
	{\mathbf{G}_{e}}_{(i,j,k+\frac{1}{2})} - {\mathbf{G}_{e}}_{(i,j,k-\frac{1}{2})} - 
	{\mathbf{G}_{v}}_{(i,j,k+\frac{1}{2})} + {\mathbf{G}_{v}}_{(i,j,k-\frac{1}{2})} 
	\right) \, \mbox{.} \nonumber
\end{eqnarray}
For the general curvilinear coordinate case 
$\Delta \xi = \Delta \eta = \Delta \zeta = 1$. The anisotropic  
artificial dissipation method of Turkel and Vatsa \cite{Turkel_Vatsa_1994}
is implemented through the modification of the inviscid flux vectors, 
$\mathbf{E}_{e}$, $\mathbf{F}_{e}$ and $\mathbf{G}_{e}$. The numerical scheme 
is nonlinear and allows the selection between artificial dissipation terms of 
second and fourth differences, which is very important for capturing discontinuities 
in the flow. The numerical fluxes are calculated at interfaces in order to reduce 
the size of the calculation cell and, therefore, facilitate the implementation 
of second derivatives since the the concept of numerical fluxes vectors is 
used for flux differencing. Only internal interfaces receive the corresponding 
artificial dissipation terms, and differences of the viscous flux vectors 
use two neighboring points of the interface. 

The inviscid flux vectors, with the addition of the artificial dissipation
contribution, can be written as
\begin{eqnarray}
	{\mathbf{E}_{e}}_{(i \pm \frac{1}{2},j,k)} 
	= \frac{1}{2} \left( {\mathbf{E}_{e}}_{(i,j,k)} + {\mathbf{E}_{e}}_{(i \pm 1,j,k)} \right)
	- J^{-1} \mathbf{d}_{(i \pm \frac{1}{2},j,k)} \, \mbox{,} \nonumber \\
	{\mathbf{F}_{e}}_{(i,j\pm \frac{1}{2},k)} 
	= \frac{1}{2} \left( {\mathbf{F}_{e}}_{(i,j,k)} + {\mathbf{F}_{e}}_{(i,j \pm 1,k)} \right)
	- J^{-1} \mathbf{d}_{(i,j \pm \frac{1}{2},k)} \, \mbox{,} \label{eq:inv_flux_vec}\\
	{\mathbf{G}_{e}}_{(i,j,k\pm \frac{1}{2})} 
	= \frac{1}{2} \left( {\mathbf{G}_{e}}_{(i,j,k)} + {\mathbf{G}_{e}}_{(i,j,k \pm 1)} \right)
	- J^{-1} \mathbf{d}_{(i,j,k \pm \frac{1}{2})} \, \mbox{,} \nonumber
\end{eqnarray}
in which the $\mathbf{d}_{(i\pm 1,j,k)}$,$\mathbf{d}_{(i,j\pm 1,k)}$ and $\mathbf{d}_{(i,j,k\pm 1)}$ terms
are the Turkel and Vatsa \cite{Turkel_Vatsa_1994} artificial dissipation terms
in the $i$, $j$, and $k$ directions respectively. The scaling of the artificial
dissipation operator in each coordinate direction is weighted by its own spectral 
radius of the corresponding flux Jacobian matrix, which gives the non-isotropic 
characteristics of the method \cite{BIGA02}. The artificial dissipation contribution
in the $\xi$ direction is given by
\begin{eqnarray}
	\mathbf{d}_{(i + \frac{1}{2},j,k)} & = & 
	\lambda_{(i + \frac{1}{2},j,k)} \left[ \epsilon_{(i + \frac{1}{2},j,k)}^{(2)}
	\left( \mathcal{W}_{(i+1,j,k)} - \mathcal{W}_{(i,j,k)} \right) \right. \label{eq:dissip_term}\\
	& & \epsilon_{(i + \frac{1}{2},j,k)}^{(4)} \left( \mathcal{W}_{(i+2,j,k)} 
	- 3 \mathcal{W}_{(i+1,j,k)} + 3 \mathcal{W}_{(i,j,k)} 
	- \mathcal{W}_{(i-1,j,k)} \right) \left. \right] \, \mbox{,} \nonumber
\end{eqnarray}
in which
\begin{eqnarray}
	\epsilon_{(i + \frac{1}{2},j,k)}^{(2)} & = &
	k^{(2)} \mbox{max} \left( \nu_{(i+1,j,k)}^{d}, 
	\nu_{(i,j,k)}^{d} \right) \, \mbox{,} \label{eq:eps_2_dissip} \\
	\epsilon_{(i + \frac{1}{2},j,k)}^{(4)} & = &
	\mbox{max} \left[ 0, k^{(4)} - \epsilon_{(i + \frac{1}{2},j,k)}^{(2)} \right] 
	\, \mbox{.} \label{eq:eps_4_dissip}
\end{eqnarray}
The original article \cite{Turkel_Vatsa_1994} recomends using $k^{(2)}=0.25$ and 
$k^{(4)}=0.016$ for the dissipation artificial constants. The pressure 
gradient sensor, $\nu_{(i,j,k)}^{d}$, for the $\xi$ direction is written as
\begin{equation}
	\nu_{(i,j,k)}^{d} = \frac{|p_{(i+1,j,k)} - 2 p_{(i,j,k)} + p_{(i-1,j,k)}|}
	                          {p_{(i+1,j,k)} - 2 p_{(i,j,k)} + p_{(i-1,j,k)}} 
	\, \mbox{.}
\label{eq:p_grad_sensor}
\end{equation}
The $\mathcal{W}$ vector from Eq.\ \eqref{eq:dissip_term} is calculated as a function of the
conserved variable vector, $\hat{Q}$, written in Eq.\ \eqref{eq:hat_Q_vec}.
The formulation intends to keep the total enthalpy constant in a final converged 
steady solution, which is the correct result for the Navier-Stokes equations with 
$Re \rightarrow \infty$. This approach is also valid for the viscous formulation 
because the dissipation terms are added to the inviscid flux terms, in which they 
are really necessary to avoid nonlinear instabilities of the numerical formulation. 
The $\mathcal{W}$ vector is given by
\begin{equation}
	\mathcal{W} = \hat{Q} + \left[0 \,\, 0 \,\, 0 \,\, 0 \,\, p \right]^{T} \, \mbox{.}
	\label{eq:W_dissip}
\end{equation}
The spectral radius-based scaling factor, $\lambda$, for the $i-\mbox{th}$ 
direction is written
\begin{equation}
	\lambda_{(i+\frac{1}{2},j,k)} = \frac{1}{2} \left[ 
	\left( \overline{\lambda_{\xi}}\right)_{(i,j,k)} + 
	\left( \overline{\lambda_{\xi}}\right)_{(i+1,j,k)}
	\right] \, \mbox{,} 
\end{equation}
where
\begin{equation}
    \overline{\lambda_{\xi}}_{(i,j,k)} = \lambda_{\xi} \left[ 1 + 
	\left(\frac{\lambda_{\eta}}{\lambda_{\xi}} \right)^{0.5} + 
	\left(\frac{\lambda_{\zeta}}{\lambda_{\xi}} \right)^{0.5} \right] 
	\, \mbox{.}
\end{equation}
The spectral radii, $\lambda_{\xi}$, $\lambda_{\eta}$ and $\lambda_{\zeta}$ are given
by
\begin{eqnarray}
	\lambda_{\xi} & = & 
	|U| + a \sqrt{\xi_{x}^{2} + \eta_{y}^{2} + \zeta_{z}^{2}} 
	\, \mbox{,} \nonumber \\
	\lambda_{\xi} & = & 
	|V| + a \sqrt{\xi_{x}^{2} + \eta_{y}^{2} + \zeta_{z}^{2}} 
	\, \mbox{,} \\
	\lambda_{\xi} & = & 
	|W| + a \sqrt{\xi_{x}^{2} + \eta_{y}^{2} + \zeta_{z}^{2}} 
	\, \mbox{,} \nonumber
\end{eqnarray}
in which, $U$, $V$ and $W$ are the contravariant velocity components in the $\xi$, $\eta$
and $\zeta$, previously written in Eq.\ \eqref{eq:vel_contrv}, and $a$ is the local 
speed of sound, which can be written as
\begin{equation}
	a = \sqrt{\frac{\gamma p}{\rho}} \, \mbox{.}
\end{equation}
The calculation of artificial dissipation terms for the other coordinate directions
are completely similar and, therefore, they are not discussed in the present work.

\subsection{Time Marching Method}

The time marching method used in the present work is a 2nd-order, 5-step Runge-Kutta
scheme based on the work of Jameson and co-workers \cite{jameson_mavriplis_86,Jameson81}. 
The time integration can be written as
\begin{equation}
	\begin{array}{ccccc}
	Q_{(i,jk,)}^{(0)} & = & Q_{(i,jk,)}^{(n)} \, \mbox{,} & & \\
	Q_{(i,jk,)}^{(l)} & = & Q_{(i,jk,)}^{(0)} -  
	& \alpha_{l} {\Delta t}_{(i,j,k)} {RHS}_{(i,j,k)}^{(l-1)} \, & 
	\,\,\,\, l = 1,2 \cdots 5, \\
	Q_{(i,jk,)}^{(n+1)} & = & Q_{(i,jk,)}^{(5)} \, \mbox{,} & &
	\end{array}
	\label{eq:localdt}
\end{equation}
in which $\Delta t$ is the time step and $n$ and $n+1$ indicate the property
values at the current and at the next time step, respectively. The literature
\cite{jameson_mavriplis_86,Jameson81} recommends 
\begin{equation}
	\begin{array}{ccccc}
		\alpha_{1} = \frac{1}{4} \,\mbox{,} & \alpha_{2} = \frac{1}{6} \,\mbox{,} &
		\alpha_{3} = \frac{3}{8} \,\mbox{,} & \alpha_{4} = \frac{1}{2} \,\mbox{,} & 
		\alpha_{5} = 1 \,\mbox{,} 
	\end{array}
\end{equation}
in order to improve the numerical stability of the time integration. The present
scheme is theoretically stable for $CFL \leq 2\sqrt{2}$, under a linear analysis
\cite{BIGA02}.
\section{Boundary Conditions} \label{sec:BC}

The present section presents all boundary conditions used for the 
turbulent compressible jet flow simulation such as inlet, outlet, 
centerline and far field boundary conditions. Moreover, the 
numerical treatment of the centerline singularity and the 
implementation of the periodic boundary in the azimuthal 
direction are also discussed in the end of the section.

\subsection{Far Field Boundary}

Riemann invariants \cite{Long91} are used to implement far field boundary conditions.
They are derived from the cha\-rac\-te\-ris\-tic relations for the Euler equations.
At the interface of the outer boundary, the following expressions apply
\begin{eqnarray}
	\mathbf{R}^{-} = {\mathbf{R}}_{\infty}^{-} & = & q_{n_\infty}-\frac{2}{\gamma-1}a_\infty\, \mbox{,}  \\
	\mathbf{R}^{+} = {\mathbf{R}}_{e}^{+} & = & q_{n_e}-\frac{2}{\gamma-1}a_e \, \mbox{,}
	\label{eq:R-farfield}
\end{eqnarray}
where $\infty$ and $e$ indexes stand for the property in the freestream and in the 
internal region, respectively. $q_n$ is the velocity component normal to the outer surface,
defined as
\begin{equation}
	q_n={\bf u} \cdot \vec{n} \, \mbox{,}
	\label{eq:qn-farfield}
\end{equation}
and $\vec{n}$ is the unit outward normal vector 
\begin{equation}
	\vec{n}=\frac{1}{\sqrt{\eta_{x}^2+\eta_{y}^2+\eta_{z}^2}}
	[\eta_x \ \eta_y \ \eta_z ]^T \, \mbox{.}
	\label{eq:norm-vec}
\end{equation}
Equation \eqref{eq:qn-farfield} assumes that the $\eta$ direction is pointing from the jet to the 
external boundary. Solving for $q_n$ and $a$, one can obtain
\begin{eqnarray}
	q_{n f} = \frac{\mathbf{R}^+ + \mathbf{R}^-}{2} \, \mbox{,} & \ & 
	a_f = \frac{\gamma-1}{4}(\mathbf{R}^+ - \mathbf{R}^-) \, \mbox{.}
	\label{eq: qn2-farfield}
\end{eqnarray}
The index $f$ is linked to the property at the boundary surface and will be used to update 
the solution at this boundary. For a subsonic exit boundary, $0<q_{n_e}/a_e<1$, the 
velocity components are derived from internal properties as
 \begin{eqnarray}
	 u_f&=&u_e+(q_{n f}-q_{n_e})\eta_x \, \mbox{,} \nonumber \\ 
	 v_f&=&v_e+(q_{n f}-q_{n_e})\eta_y \, \mbox{,} \\ 
	 w_f&=&w_e+(q_{n f}-q_{n_e})\eta_z \, \mbox{.} \nonumber
	 \label{eq:vel-farfield}
 \end{eqnarray}
Density and pressure properties are obtained by extrapolating the entropy from 
the adjacent grid node,
\begin{eqnarray}
	\rho_f = 
	\left(\frac{\rho_{e}^{\gamma}a_{f}^2}{\gamma p_e} \right)^{\frac{1}{\gamma-1}}
	\, \mbox{,} & \ &
	p_{f} = \frac{\rho_{f} a_{f}^2}{\gamma} \, \mbox{.} \nonumber
	 \label{eq:rhop-farfield}
\end{eqnarray}
For a subsonic entrance, $-1<q_{n_e}/a_e<0$, properties are obtained similarly 
from the freestream variables as
\begin{eqnarray}
	u_f&=&u_\infty+(q_{n f}-q_{n_\infty})\eta_x \, \mbox{,} \nonumber \\
	v_f&=&v_\infty+(q_{n f}-q_{n_\infty})\eta_y \, \mbox{,} \\
	w_f&=&w_\infty+(q_{n f}-q_{n_\infty})\eta_z \, \mbox{,} \nonumber
	\label{eq:vel2-farfield}
\end{eqnarray}
\begin{equation}
	\rho_f = 
	\left(\frac{\rho_{\infty}^{\gamma}a_{f}^2}{\gamma p_\infty} \right)^{\frac{1}{\gamma-1}}
	\, \mbox{.}
	\label{eq:rhop2-farfield}
\end{equation}
For a supersonic exit boundary, $q_{n_e}/a_e>1$, the properties are extrapolated 
from the interior of the domain as
\begin{eqnarray}
	\rho_f&=&\rho_e \, \mbox{,} \nonumber\\
	u_f&=&u_e \, \mbox{,} \nonumber\\
	v_f&=&v_e \, \mbox{,} \\
	w_f&=&w_e \, \mbox{,} \nonumber\\
	e_f&=&e_e \, \mbox{,} \nonumber   
	\label{eq:supso-farfield}
\end{eqnarray}
and for a supersonic entrance, $q_{n_e}/a_e<-1$, the properties are extrapolated 
from the freestream variables as
\begin{eqnarray}
	\rho_f&=&\rho_\infty \, \mbox{,}  \nonumber\\
	u_f&=&u_\infty \, \mbox{,}  \nonumber\\
	v_f&=&v_\infty \, \mbox{,} \\
	w_f&=&w_\infty \, \mbox{,} \nonumber\\
	e_f&=&e_\infty \, \mbox{.} \nonumber
	\label{eq:supso2-farfield}
\end{eqnarray}

\subsection{Entrance Boundary}

For a jet-like configuration, the entrance boundary is divided in two areas: the
jet and the area above it. The jet entrance boundary condition is implemented through 
the use of the 1-D characteristic relations for the 3-D Euler equations for a flat
velocity profile. The set of properties then determined is computed from within and 
from outside the computational domain. For the subsonic entrance, the $v$ and $w$ components
of the velocity are extrapolated by a zero-order extrapolation from inside the 
computational domain and the angle of flow entrance is assumed fixed. The rest of the properties 
are obtained as a function of the jet Mach number, which is a known variable. 
\begin{eqnarray}
	\left( u \right)_{1,j,k} & = & u_{j} \, \mbox{,} \nonumber \\
	\left( v \right)_{1,j,k} & = & \left( v \right)_{2,j,k} \,\mbox{,} \\
	\left( w \right)_{1,j,k} & = & \left( w \right)_{2,j,k} \, \mbox{.} \nonumber
	\label{eq:vel-entry}
\end{eqnarray}
The dimensionless total temperature and total pressure are defined with the isentropic relations:
\begin{eqnarray}
	T_t = 1+\frac{1}{2}(\gamma-1)M_{\infty}^{2} \, & \mbox{and} & 
	P_t = \frac{1}{\gamma}(T_t)^{\gamma / (\gamma-1)} \, \mbox{.}
	\label{eq:Tot-entry}
\end{eqnarray}
The dimensionless static temperature and pressure are deduced from Eq.\ \eqref{eq:Tot-entry},
resulting in
\begin{eqnarray}
	\left( T \right)_{1,j,k}=\frac{T_t}{1+\frac{1}{2}(\gamma-1)(u^2+v^2+w^2)_{1,j,k}} \, 
	& \mbox{and} & 
	\left( p \right)_{1,j,k}=\frac{1}{\gamma}(T)_{1,j,k}^{\gamma / (\gamma-1)} \, \mbox{.}
	\label{eq:Stat-entry}
\end{eqnarray}
For the supersonic case, all conserved variables receive jet property values.

The far field boundary conditions are implemented outside of the jet area in order to correctly
propagate information comming from the inner domain of the flow to the outter region of 
the simulation. However, in the present case, $\xi$, instead of $\eta$, as presented in 
the previous subsection, is the normal direction used to define the Riemann invariants.

\subsection{Exit Boundary Condition}

At the exit plane, the same reasoning of the jet entrance boundary is applied. This time, 
for a subsonic exit, the pressure is obtained from the outside and all other variables are 
extrapolated from the interior of the computational domain by a zero-order extrapolation. The 
conserved variables are obtained as
\begin{eqnarray}
	(\rho)_{I_{MAX},j,k} &=& \frac{(p)_{I_{MAX},j,k}}{(\gamma-1)(e)_{I_{MAX}-1,j,k}} \mbox{,} \\
	(\vec{u})_{I_{MAX},j,k} &=& (\vec{u})_{I_{MAX}-1,j,k}\mbox{,} \\
	(e_i)_{I_{MAX},j,k} &=& 
	(\rho)_{I_{MAX},j,k}\left[ (e)_{I_{MAX}-1,j,k}+
	\frac{1}{2}(\vec{u})_{I_{MAX},j,k}\cdot(\vec{u})_{I_{MAX},j,k} \right] \, \mbox{,}
	\label{eq:exit}
\end{eqnarray}
in which $I_{MAX}$ stands for the last point of the mesh in the axial direction. For 
the supersonic exit, all properties are extrapolated from the interior domain.

\subsection{Centerline Boundary Condition}

The centerline boundary is a singularity of the coordinate transformation, and, hence, 
an adequate treatment of this boundary must be provided. The conserved properties 
are extrapolated from the ajacent longitudinal plane and are averaged in the azimuthal 
direction in order to define the updated properties at the centerline of the jet.

\begin{figure}[htb!]
       \begin{center}
       {\includegraphics[width=0.35\textwidth]{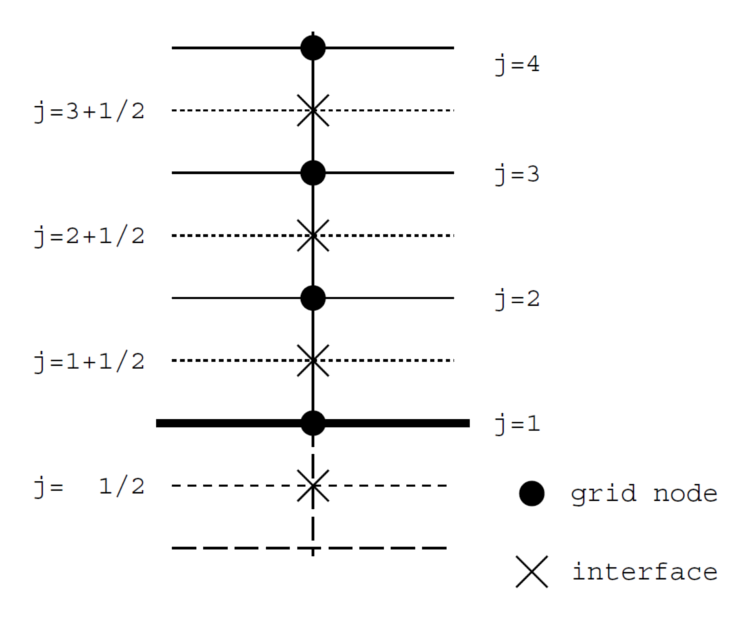}}\\
	   \caption{Boundary point distribution in the calculation of dissipation operator at the 
	            centerline \cite{BIGA02}.}
	   \label{fig:centerline}
       \end{center}
\end{figure}

The fourth-difference terms of the artificial dissipation scheme, used in the present 
work, are carefully treated in order to avoid the five-point difference stencils at 
the centerline singularity. 
If one considers the flux balance at one grid point near the centerline boundary in 
a certain coordinate direction, let $w_{j}$ denote a component of the $\mathcal{W}$ 
vector from Eq.\ \eqref{eq:W_dissip} and $\mathbf{d}_{j}$ denote the corresponding artificial
dissipation term at the mesh point $j$. In the present example, 
$\left(\Delta w\right)_{j+\frac{1}{2}}$ stands for the difference between the solution
at the interface for the points $j+1$ and $j$. The fouth-difference of the dissipative
fluxes from Eq.\ \eqref{eq:dissip_term} can be written as
\begin{equation}
	\mathbf{d}_{j+\frac{1}{2}} = \left( \Delta w \right)_{j+\frac{3}{2}} 
	- 2 \left( \Delta w \right)_{j+\frac{1}{2}}
	+ \left( \Delta w \right)_{j-\frac{1}{2}} \, \mbox{.}
\end{equation}
Considering the centerline and the point $j=1$, as presented in 
Fig.\ \ref{fig:centerline}, the calculation of $\mathbf{d}_{1+\frac{1}{2}}$ demands the 
$\left( \Delta w \right)_{\frac{1}{2}}$ term, which is unknown since it is outside the
computation domain. 
In the present work a extrapolation is performed and given by
\begin{equation}
	\left( \Delta w \right)_{\frac{1}{2}} =
	- \left( \Delta w \right)_{1+\frac{1}{2}} \, \mbox{.}
\end{equation}
This extrapolation modifies the calculation of $\mathbf{d}_{1+\frac{1}{2}}$ that can be written as
\begin{equation}
	\mathbf{d}_{j+\frac{1}{2}} = \left( \Delta w \right)_{j+\frac{3}{2}} 
	- 3 \left( \Delta w \right)_{j+\frac{1}{2}} \, \mbox{.}
\end{equation}
The approach is plausible since the centerline region is smooth and it does not have high
gradients of properties.

\subsection{Periodic Boundary Condition}

A periodic condition is implemented between the first ($K=1$) and the last point in the 
azimutal direction ($K=K_{MAX}$) in order to close the 3-D computational domain. There 
are no boundaries in this direction, since all the points are inside the domain. The first
and the last points, in the azimuthal direction, are superposed in order to facilitate
the boundary condition implementation which is given by
\begin{eqnarray}
	(\rho)_{i,j,K_{MAX}} &=& (\rho)_{i,j,1} \, \mbox{,} \nonumber\\
	(u)_{i,j,K_{MAX}} &=& (u)_{i,j,1} \, \mbox{,} \nonumber\\
	(v)_{i,j,K_{MAX}} &=& (v)_{i,j,1} \, \mbox{,} \\
	(w)_{i,j,K_{MAX}} &=& (w)_{i,j,1} \, \mbox{,} \nonumber\\
	(e)_{i,j,K_{MAX}} &=& (e)_{i,j,1} \, \mbox{.} \nonumber
	\label{eq:periodicity}
\end{eqnarray}



\section{Dynamic Mode Decomposition}

\subsection{Theoretical Framework}

The DMD method provides a spatio-temporal decomposition of the flow into a set of
dynamic modes that are derived from time-resolved snapshots. 
For example, a generic flow variable, ${\bf x}_{\text{DMD}}(x,y,z,t)$, where $x$, $y$, 
$z$ and $t$ stand for spatial coordinates and time, respectively, can be 
represented by
\begin{equation}
	{\bf x}_{\text{DMD}}(x,y,z,t) = \sum^{m-1}_{i=1} a_i \, \exp(\lambda_i t) \, \phi_i(x,y,z) \mbox{ .}
  \label{eq:dmd0}
\end{equation}
Here, $a_i$ and $\lambda_i$ are the amplitude and the frequency of the spatial mode
$\phi_i$. The underlying mathematics is closely related to the idea of the Arnoldi
algorithm \cite{schmid2010dynamic}. This flow variable, extracted from the simulation,
can be represented in the form of a snapshot sequence $ {\bf X} = \left[{\bf x}(t_{1}) \; {\bf x}(t_{2}) \; 
\cdots\; {\bf x}(t_{m}) \right] \,\, \in \,\, \mathbb{R}^{n\times m}$, where ${\bf x}(t_{i}) \,\, \in 
\,\, \mathbb{R}^{n}$ is the $i$-th snapshot, $m$ denotes the number of snapshots
and $n$, the spatial dimension of each time snapshot. Each snapshot, ${\bf x}(t_{i})$, contains
a set of variables depending on the user's choice. The present study is designed to 
collect data regularly separated in time by $\Delta t$ even though recent techniques 
allow irregularly spaced sampling in time of the data \cite{tu2014spectral}. The authors assume 
that there exists a linear operator $\mathcal{A} \,\, \in \,\, \mathbb{R}^{n\times n}$ 
connecting two consecutive snapshot giving
\begin{equation}
  {\bf x}_{i+1} = \mathcal{A} \, {\bf x}_i \;\;\;\;\;  \text{for}\;\; i = 1,\cdots,\text{m}-1 \mbox{.}
  \label{eq:dmd1}
\end{equation}
The $\mathcal{A}$ operator is an approximation of the Koopman operator
\cite{rowley2009spectral}, whose eigen-elements can approximate the
underlying dynamics of the flow, even if such dynamics is nonlinear. The objective of
the DMD is the determination of these characteristics. The selection of the eigen-elements 
of $\mathcal{A}$ is a matter of importance since the accuracy of the results, as well as
the computational costs, both depend significantly on the method of choice.
The strategy used in the present work
is a combination of the total least-squares DMD, described in Ref.\ \citen{hemati2015biasing},
and the streaming DMD algorithm presented in Hemati {\it et al.}
\cite{hemati2014dynamic}. The former technique provides a noise-aware DMD
technique while the latter allows the assimilation ``on-the-fly'' of new
incoming snapshots and it can even theoretically include an infinite number, $m$,
of snapshots. Hemati { et al.} \cite{hemati2016improving} ran successfully this
combined technique to analyze the dynamics of the flow separation over a flat
plate. In practice, the $\mathcal{A}$ DMD operator of Eq.\ \eqref{eq:dmd1} can be defined
as $\mathcal{A} = {\bf Y} {\bf X}^{+}$ using the previously defined snapshot matrix, ${\bf X}$, and 
its time-shifted version ${\bf Y} = \left[{\bf x}(t_{1}+\Delta t) \; {\bf x}(t_{2}+\Delta t) \; \cdots \;
{\bf x}(t_{m}+\Delta t) \right]$. In this relation, ${\bf X}^+$ stands for the
Moore-Penrose pseudoinverse of ${\bf X}$. The solution of the problem in the present form is 
prohibitively expensive in terms of CPU and memory costs. The streaming DMD approach suggests a
solution to reformulate $\mathcal{A}$ in order to be able to handle large dimension problems.
First, the augmented snapshot matrix, ${\bf Z} = [{\bf X} \, {\bf Y}]^T$, is built\cite{hemati2015biasing}.
After substitution, a low-dimensional version of $\mathcal{A}$, $\tilde{\mathcal{A}}$, can be obtained under the form
\begin{equation}
\tilde{\mathcal{A}} \, = \, {\bf Q}^{T}_{x} \, \left[ \begin{array}{cc} {\bf 0} & {\bf I} \end{array} \right] \,
{\bf Q}_{z} \, {\bf G}_{z} \, {\bf Q}^{T}_{z} \, \left[ \begin{array}{c} {\bf 0} \\ {\bf I} \end{array} \right] 
\, {\bf Q}_{x} {\bf G}^{+}_{x} \,\,\, \in \,\, \mathbb{R}^{r \times r} \mbox{ ,}
\label{eq:Atilde}
\end{equation}
where $r$ is the rank of ${\bf X}$, ${\bf Q}_{x}$ and ${\bf Q}_{z}$ are obtained from the QR-decomposition of 
${\bf X}$ and ${\bf Z}$, respectively. Hence, one could write that ${\bf X} = {\bf Q}_{x} {\bf R}_{x}$ and 
${\bf Z} = {\bf Q}_{z} {\bf R}_{z}$. Therefore, one can also write that ${\bf G}_{z} = {\bf R}_{z} {\bf R}^{T}_{z}$ 
and ${\bf G}_{x} = {\bf R}_{x} {\bf G}_{z} {\bf R}^{T}_{z}$. This procedure allows an incremental update 
of new available snapshots, without storing all of them in memory.
Moreover, in this expression, the total number snapshot, $m$, does not appear anymore. During the 
streaming DMD process, a POD compression is included allowing the user to choose the rank of the DMD operator, $r$.
The DMD modes and frequencies are given by the eigenvectors and eigenvalues of $\tilde{\mathcal{A}}$,
such that $\phi_{i}$ is the $i$-th eigenvector with the associated eigenvalue, $\mu_{i}$. Hence, 
the associated growth rate and frequency of the $i$-th DMD mode are given by
\begin{equation}
	\sigma_{i} = \frac{\mbox{log}(|\mu_{i}|)}{\Delta t} \;\;\;\; 
	\mbox{and} \;\;\;\; \omega_{i} = \frac{\mbox{arg}(\mu_{i})}{\Delta t} \mbox{ .}
\end{equation}
Finally, $\lambda_{i} = \log (\mu_{i}) / \Delta t$. 
Another interesting aspect of the DMD, is that knowing the first snapshot and the eigenvalues of the
DMD operator, one can predict the temporal behavior of the mode.
Indeed, using a discretized version of Eq. (\ref{eq:dmd0}) expressed at any time instant $k=1,\cdots,m-1$, 
\begin{equation}
  {\bf x}_k = \sum^{m-1}_{i=1} \theta_{i} (k) \, \phi_{i} \mbox{,}
\end{equation}
where $\theta_i$ are the temporal coefficients of the eigenvectors $\phi_{i}$.
It comes directly, using Eq.\ (\ref{eq:dmd1}), that
\begin{equation}
	\begin{array}{ccccccc}
  {\bf x}_{k+1} &=& \mathcal{A} \, {\bf x}_{k} 
                &=& \sum\limits^{m-1}_{i=1}\theta_i(k)\mathcal{A} \, {\bf \phi}_i
				&=& \sum\limits^{m-1}_{i=1}\theta_i(k)\mu_i{\bf \phi}_i\\\\
                &=& \mathcal{A}^k \, {\bf x}_{1} 
				&=& \sum\limits^{m-1}_{i=1}\theta_i(1)\mu^k_i{\bf \phi}_i \mbox{,}
    \end{array}
    \label{eq:dmd3}
\end{equation}
Using the work of Ref.\ \citen{kutz2016multiresolution}, the matrix of the initial
coefficient can be calculated using the relation
\begin{equation}
	\boldsymbol{\theta}(1) = \boldsymbol{\phi}^+ \boldsymbol{x}_1 \mbox{.}
	\label{eq:theta1}
\end{equation}

\subsection{Choice of the Parameters}

Two main parameters are considered in the DMD framework initially introduced by Schmid.\cite{schmid2010dynamic}
The first one is $\Delta t$, the constant time-step between two consecutive snapshots,
while the second one is $m$, the total number of snapshots.
Both of them require a good knowledge of the physical phenomenon under study.
According to Schmid \cite{schmid2010dynamic}, the sample rate must be sufficiently high, 
about three times the Nyquist cutoff, to capture correctly the dynamics of an oscillatory flow.
The idea is, then, to tune the sampling frequency based on the phenomenon the user wants to study.
However, following Chen {\it et al.} \cite{chen2012variants}, using a high sample rate,
the snapshots are likely to be correlated in time. 
This is a problem since the method impose the use of a linear independent dataset to work properly.
Finally, a high number, $m$, of snapshots could also affect the linear independency of the snapshots.
In the algorithm used in the present work, a Gram-Schmidt step is included in the process to address this 
problem\cite{hemati2014dynamic,tissot2013dynamic}.

  \section{Case of a High Reynolds Number Supersonic Jet Flow}\label{sec:LES}

The present section is devoted to the study of a supersonic 
perfectly expanded jet flow. The geometry and flow configurations 
of interest are presented, followed by large eddy simulation results, 
which are compared to analytical, numerical and experimental data from 
the literature \cite{mendez2010large,bridges2008turbulence}. The LES 
results provide a database for three DMD studies using the velocity 
magnitude, the vorticity, represented by the Q criterion, and the 
divergence of the flow velocity.

\subsection{Geometry and Mesh Configurations}

\begin{figure}[htb!]
	\centering
    {\includegraphics[trim=1.0mm 0.5mm 1.5mm 1.0mm, clip, width=0.6\textwidth]{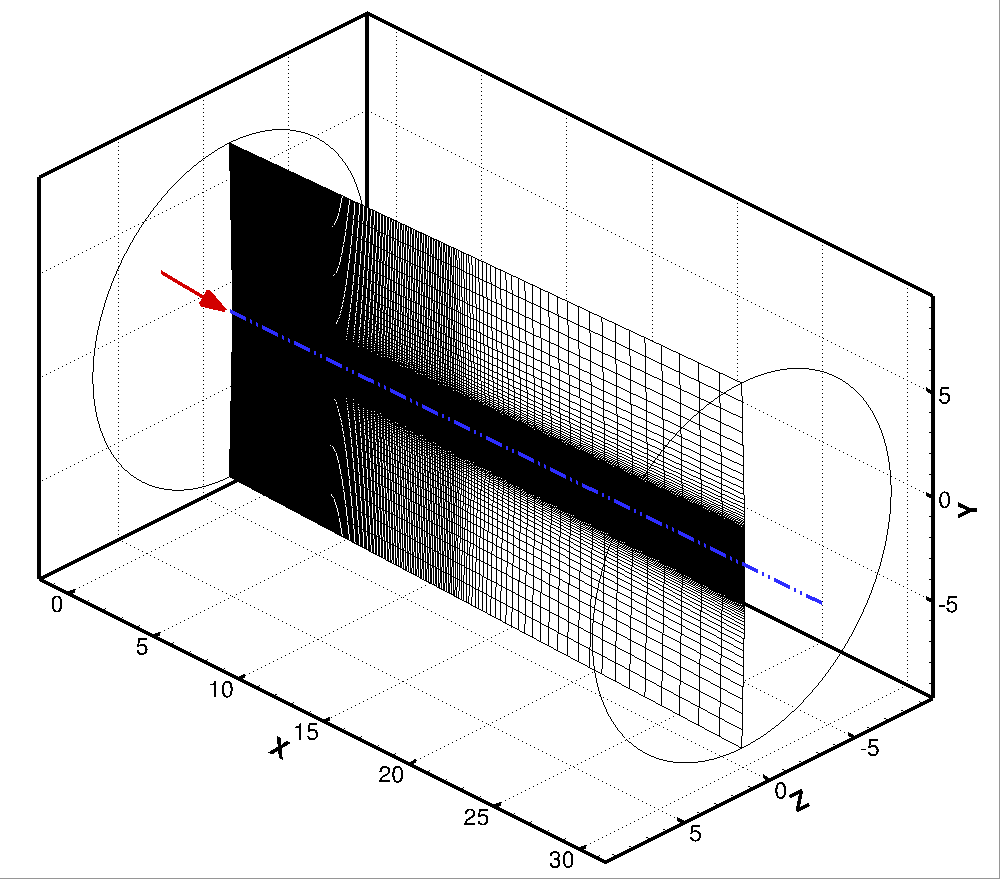}}
	\caption{3-D view of two XY slices of the grid, located above and below the centerline highlighted by a discontinuous blue line.
	The red arrow indicates the jet entrance inside the domain.}
	\label{fig:domain-mesh}
\end{figure}

Figure \ref{fig:domain-mesh} illustrates a three-dimensional view of
the representative domain for the jet flow simulations. The geometry 
resembles a frustrum of a cone with the jet entering the computational 
domain through the small base at $x=0$, and leaving the domain at the 
large base at $x=30D$. The radii of the entrance and exit plans are 
approximately 8D and 9D, respectively. The authors have chosen not to 
include the nozzle geometry in the computational domain. Hence, 
the jet entrance is located at $x=0$, for $|r|/D \leq 0.5$, where
$|r|=\sqrt{y^2+z^2}$ is the distance from the centerline in the radial 
direction and $D$ is the incoming jet diameter. The computational domain
is created in two steps. First, a 2-D region is generated. In the
sequence, this region is rotated around the horizontal direction, $x$,
indicated by the discontinuous blue line in Fig.\ \ref{fig:domain-mesh},
in order to generate a fully 3-D geometry. The rotation approach generates
a singularity at the centerline of the domain. The treatment of this 
region is discussed in the boundary conditions section. 

The commercial mesh generator ANSYS\textsuperscript{\textregistered} ICEM CFD 
\cite{ICEM} is used for the creation of the 2-D domain for an azimuthal plane.
The zones of this geometry are created based on results from simulations of previous work \cite{Junior16}
in order to refine the mesh in the shear layer region of the flow until $x=10D$,
after the end of the potential core.
The mesh is, then, coarsened towards the outer regions of the domain 
in order to dissipate properties of the flow far from the jet.
Such mesh refinement approach can avoid reflection of information into the domain.
The radial and longitudinal dimensions of the smallest distance between mesh points of the computational grid are given
by $(\Delta \underline{r})_{min}=0.002$ and $(\Delta \underline{x})_{min}=0.0126$, respectively.
This minimal spacing occurs at the lipline of the jet and at the entrance of the computational domain. 
These dimensions are based on a reference grid of Mendez {\it et al.} \cite{mendez2010large,mendez2012large}.
The resulting computational grid is composed by 537 points in the axial direction, 442 points in the radial direction 
and 360 points in the azimuthal direction, yielding approximately 85 million grid points.
For further details about the mesh generation, the reader is referred to the work of Junqueira-Junior\cite{Junior16PhD}.

\subsection{Flow Configuration}

The flow is characterized by an 
unheated perfectly expanded jet with a Mach number of 
$1.4$ at the domain entrance. Therefore, the pressure ratio, 
$PR = P_{j}/P_\infty$, and the temperature ratio, $TR=T_{j}/T_\infty$, 
between the jet exit and the ambient freestream are equal to one, 
$PR = 1$ and $TR=1$. The time step used in the simulation is constant 
and equal to $2.0 \times 10^{-4}$ in dimensionless form. The Reynolds
number of the jet is $Re = 1.57 \times 10^{6}$, based on the jet
entrance diameter. This flow configuration is
chosen due to the absence of strong shocks waves. Strong 
discontinuities must be carefully treated using numerical approaches, 
and the authors did not want to deal with those issues at the present time. 
Moreover, numerical and experimental data for a perfectly expanded jet flow 
configuration, such as the one used in the present work, are available in the 
literature such as the work of Mendez {\it et al.} \cite{mendez2010large, mendez2012large} 
and the work of Bridges and Wernet \cite{bridges2008turbulence}.

Properties of flow at the inlet and at the far field regions have to be 
provided to the code in order to impose the boundary conditions. Density, 
$\rho$, temperature, $T$, velocity, $U$, Reynolds number, $Re$, and specific 
heat at constant volume, $C_{v}$, are provided in the dimensionless form 
to the simulation. These dimensionless properties are given by
\begin{eqnarray}
	\rho_{j} = 1.00 \, \mbox{ ,} & 
	\rho_{\infty} = 1.00 \, \mbox{ ,} \nonumber \\
	T_{j} = 1.00 \, \mbox{ ,} & 
	T_{\infty} = 1.00 \, \mbox{ ,} \\
	U_{j} = 1.4 \, \mbox{ ,} & 
	U_{\infty} = 0.00 \nonumber \, \mbox{ ,}\\
	Re_{j}=1.57 \times 10^{6} \, \mbox{ ,}& 
	C_{v} = 1.786 \, \mbox{ ,} \nonumber
\end{eqnarray}
where the $j$ subscript stands for property at the jet entrance and the 
$\infty$ subscript stands for property at the far field region. 

\subsection{Data Extraction Procedure}

For the present study, data are extracted after a preliminary simulation is run
in order to achieve a statistically steady state condition for the jet flow.
This initial preliminary simulation lasts 96 dimensionless time units.
For the current jet exit Mach number of $M_{j}=1.4$, 
this simulation time represents approximately 3 flow-through times (FTT).
One flow-through time is the time for a particle 
to cross the entire domain from the jet entrance to the domain exit.
After the flow initialization process, 
the simulations are restarted and run for another period of time in which 
data of the flow are extracted and recorded at a fixed frequency. 

\begin{table}[htbp]
\begin{center}
	\caption{Data extraction characteristics}
\label{tab:simu}
\begin{tabular}{|c|c|c|c|c|c|}
\hline\hline
Simulation & $\Delta t \, c_\infty /D$ &  No.\ Extractions & Grid Size & Total Time & FTT \\
\hline
LES statistics & 0.06 & 4096  & $500 \times 425$ (2-D) & 245.76 &  $\approx 8$ \\
DMD & 0.12 &  256  & $473 \times 412 \times 180$ (3-D) &  30.72 &  $\approx 1$ \\
\hline\hline
\end{tabular}
\end{center}
\end{table}

The temporal characteristics of the data extraction are displayed in Tab.\ \ref{tab:simu}
for both the LES statistics and the DMD computations. These data processing methods 
are very different one from each other, especially because of the grid dimension. In 
the present work, the LES statistics are computed only along 2-D surfaces, whereas 
DMD calculations use three-dimensional snapshots as input. The snapshots extracted 
during the DMD process have more than 35 million points and they are stored in the 
PLOT3D format\footnote{https://www.grc.nasa.gov/www/wind/valid/plot3d.html}, adapted for structured meshes. 
The memory size of one snapshot, used for 
the DMD calculations, is about 1.5 Gb. On the other hand, the time-dependent LES surfaces 
are all included in one single CGNS file of 40 Gb. Finally, the total simulation time, 
necessary for obtaining the LES statistics, is higher than that used in Ref.\ \citen{bres2012towards}
for the same purpose, and this can be considered as a fairly large time sample for an LES 
calculation. As indicated in Tab.\ \ref{tab:simu}, a total of 8 flow-through times have been 
used in order to obtain the LES statistics.


\subsection{Large Eddy Simulation Results}


In this subsection, 2-D distributions of properties and profiles 
are collected from the compressible LES simulation and 
compared with numerical and experimental results from the 
literature \cite{mendez2010large,mendez2012large,bridges2008turbulence}. 
\begin{figure}[htb!]
  \centering
  \subfigure[Time averaged axial velocity component.]
     {\includegraphics[trim=1.0mm 0.5mm 1.5mm 1.0mm, clip,width=0.75\textwidth]{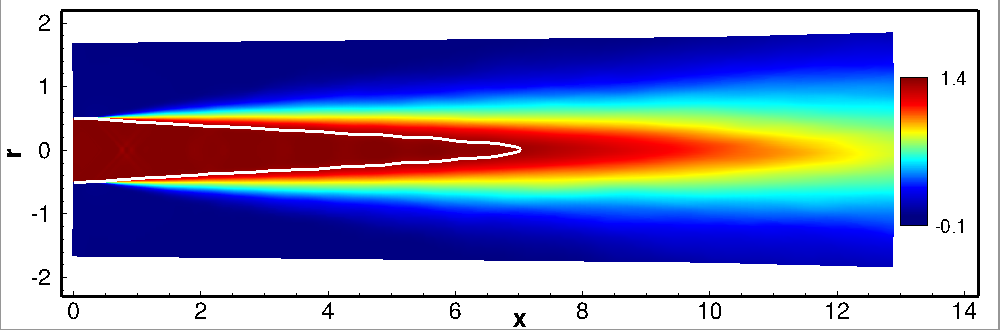}}
  \subfigure[RMS value of the fluctuating part of the axial velocity component.]
     {\includegraphics[trim=1.0mm 0.5mm 1.5mm 1.0mm, clip,width=0.75\textwidth]{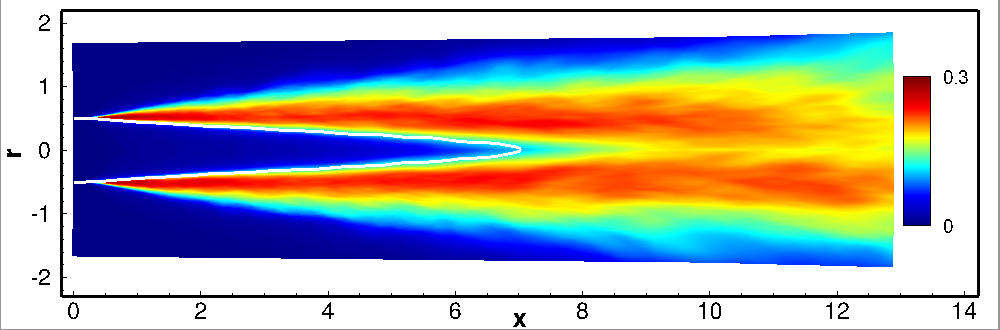}}
  \subfigure[Turbulent kinetic energy.]
     {\includegraphics[trim=1.0mm 0.5mm 1.5mm 1.0mm, clip,width=0.75\textwidth]{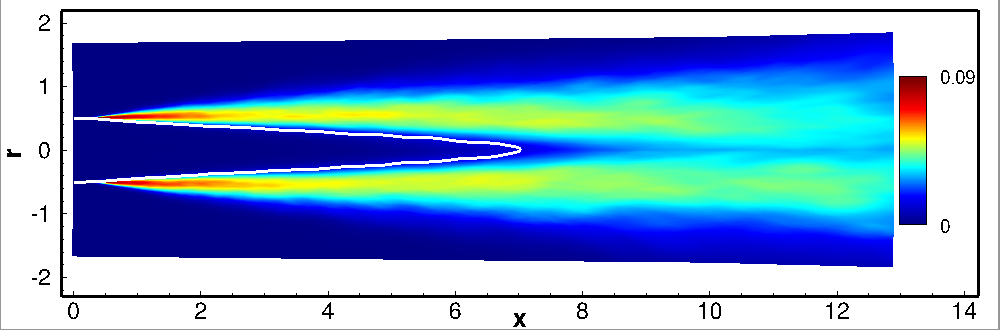}}
	\caption{Contour plots of longitudinal planes of statistically converged jet properties. The white line defines the potential core of the jet, where $u = 0.95 U_{j}$.}
	\label{fig:lat-u-av}
\end{figure}
A longitudinal plane view of the statistically-converged time-averaged distributions of three flow 
properties, namely, axial velocity component, $\langle U \rangle$, RMS value of the fluctuating part 
of the axial velocity component,  $\text{u}_{rms}$, and turbulent kinetic energy, $k$, are presented 
in Fig.\ \ref{fig:lat-u-av}. The statistical properties of the LES results are calculated using much 
more snapshots and with a more refined time increment than the numerical reference 
data \cite{mendez2010large,mendez2012large}. Each variable displays a fairly smooth flow field, 
confirming the good statistical convergence of the results. Moreover, the contours of 
$\langle U \rangle$, $\text{u}_{rms}$ and $k$ display a classical shape, with $\text{u}_{rms}$ spreading 
along with the jet shear layer and with high values of $k$ at the beginning of the mixing layer.
The white solid line defines the jet potential core region, $U_{j}^{95\%}$, which is a characteristic 
parameter of jet flows. The potential core length, $\delta_{j}^{95\%}$, is defined as the distance 
from the jet entrance and along the centerline until the jet velocity reaches $95\%$ of the velocity of 
the jet at the inlet.

In line with previous work, reported in Ref.\ \citen{Junior16}, the current simulation aims to 
reduce the error with respect to the experimental data in Ref.\ \cite{bridges2008turbulence} by 
refining the grid in the jet potential core. Table \ref{tab:core-mesh} presents the size of 
the potential cores for the current simulation, compared to the numerical results in Refs.\ \citen{Junior16},
\citen{mendez2010large} and \citen{mendez2012large}. The table presents the relative error compared to the 
experimental data \cite{bridges2008turbulence}.
The present LES calculations are performed on the same grid geometry used in Ref.\ \citen{Junior16}, 
but with more points inside the potential core. As one can see in the table, the error has been reduced 
from $26\%$ to $22\%$. The grid used in the present work needs to be further refined in order to overcome 
the dissipative characteristics of 2nd-order scheme used and, hence, keep reducing the magnitude of error 
when compared to experimental data.

\begin{table}[htbp]
\begin{center}
  \caption{Potential core length comparisons.}
  \label{tab:core-mesh}
  \begin{tabular}{|c|c|c|}
  \hline\hline
  Simulation & $\delta_{j}^{95\%}$ & Relative error\\
  \hline
  Current work & 7.05 & 22\%\\
  Junqueira-Junior {\it et al.} \cite{Junior16} & 6.84 & 26\%\\
  Mendez {\it et al.} \cite{mendez2010large,mendez2012large} & 8.35 & 8\%\\
  \hline\hline
  \end{tabular}
\end{center}
\end{table}

\begin{figure}[htb!]
  \centering
  \subfigure[Centerline]
    {\includegraphics[trim=1.0mm 1.0mm 1.5mm 1.0mm, clip, width=0.48\textwidth]{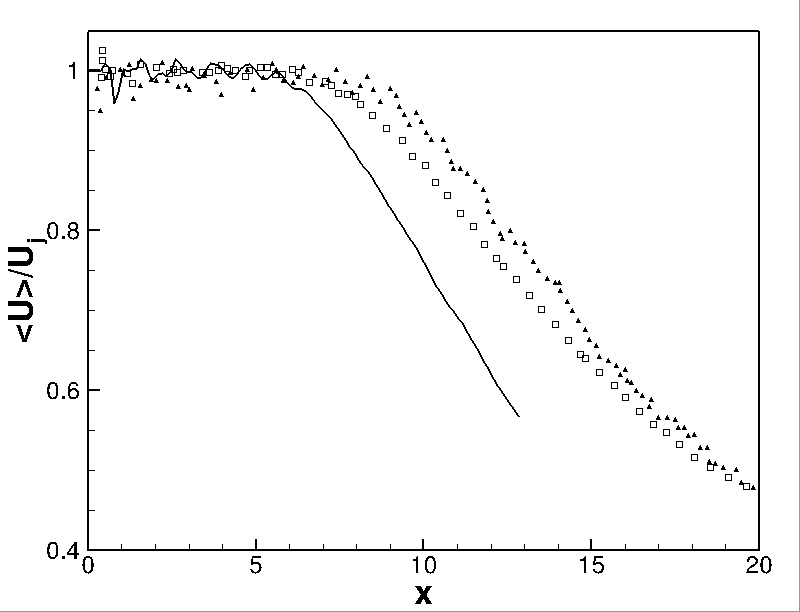}}
  \subfigure[Lipline]
    {\includegraphics[trim=1.0mm 1.0mm 1.5mm 1.0mm, clip,width=0.48\textwidth]{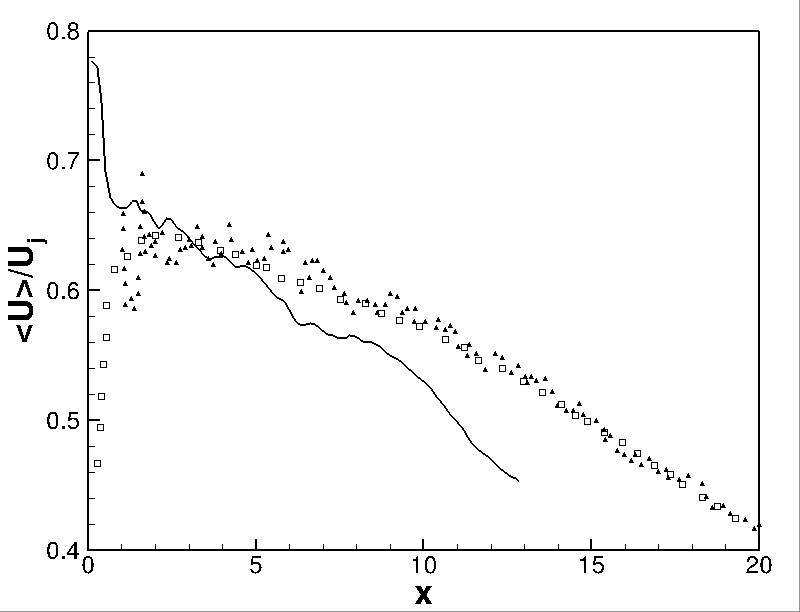}}
	\caption{Averaged axial component of velocity along the centerline and lipline.
    The solid line stands for the results of the present case, the open square symbols 
    represent the numerical references \cite{mendez2010large,mendez2012large} and the 
	triangular symbols are the experimental data \cite{bridges2008turbulence}.}
	\label{fig:u-av}
\end{figure}

\begin{figure}[htb!]
  \centering
  \subfigure[Centerline]
    {\includegraphics[trim=1.0mm 1.0mm 1.5mm 1.0mm, clip,width=0.48\textwidth]{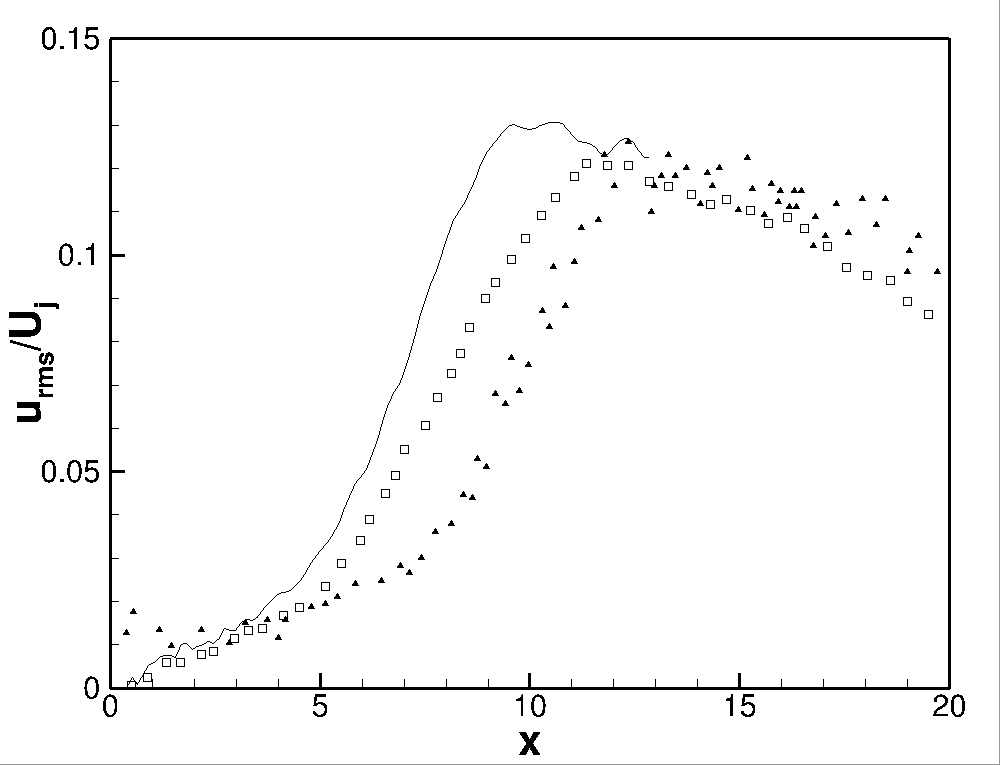}}
  \subfigure[Lipline]
    {\includegraphics[trim=1.0mm 1.0mm 1.5mm 1.0mm, clip,width=0.48\textwidth]{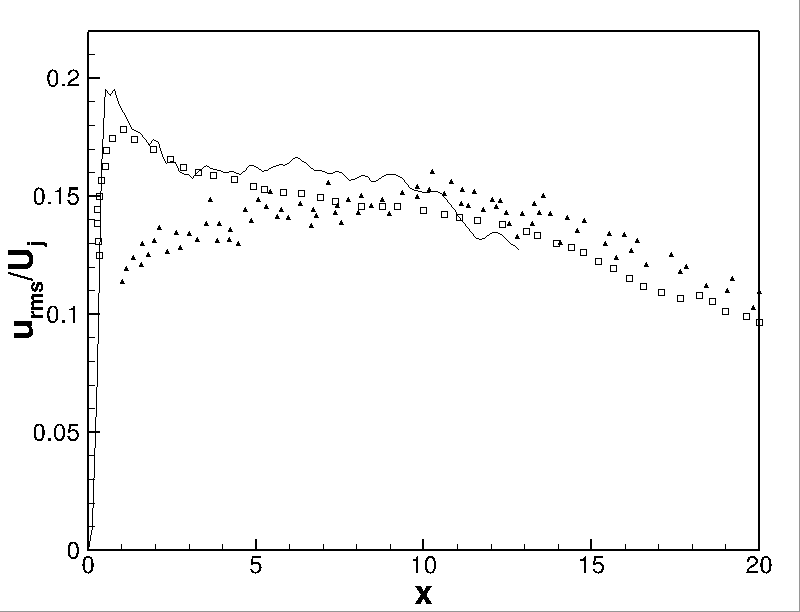}}
	\caption{RMS value of the fluctuating part of the axial component of velocity along the centerline and lipline.
    The solid line stands for the results of the present case, the open square symbols 
    represent the numerical references \cite{mendez2010large,mendez2012large} and the 
	triangular symbols are the experimental data \cite{bridges2008turbulence}.}
	\label{fig:u-rms}
\end{figure}

The evolution of the averaged axial component of velocity 
and the evolution of the RMS value of the fluctuating part of the axial component of velocity 
along the centerline and the lipline are illustrated in 
Figs.\ \ref{fig:u-av} and \ref{fig:u-rms}, respectively. 
The solid line stands for the results of the present case, 
the open square symbols represent the LES results of 
Mendez {\it et al.} \cite{mendez2010large,mendez2012large}, 
while the triangular symbols stand for the experimental 
data of Bridges and Wernet \cite{bridges2008turbulence}. 
The lipline is the surface defined over $r=0.5D$, which
represents the boundary of the jet at the entrance of the 
domain. The comparison of profiles indicates that 
distributions of $\langle U \rangle/U_{j}$ along the 
centerline correlates well with the references until 
$x = 7.0 D$, where the grid has good resolution. 
The time averaged axial component of velocity start 
to correlate poorly with the reference when the mesh 
spacing increases, $x > 7.0 D$, due to the mesh coarsening 
in the streamwise direction. The mesh coarsening is used 
in order to add artificial dissipation towards the exit of 
the domain, since the numerical framework does not have
a sponge zone implemented. The time averaged axial 
component of velocity calculated along the lipline 
correlates well with the references until $x \approx 6.0 D$. 
The magnitude of $\langle U \rangle/U_{j}$ along the 
lipline is understimated for $x > 6.0 D$.

The distribution of $u_{rms}/U_{j}$ calculated along the 
centerline fits the numerical and experimental reference 
distributions of the same property for $x < 4.0 D$. However,
it presents an overestimated distribution of $u_{rms}/U_{j}$
when compared with both numerical and experimental data 
at other positions along the centerline. The numerical 
reference has also calculated an overestimated distribution 
of $u_{rms}/U_{j}$along the centerline when compared to 
the experimental reference at $x > 5.0 D$. The distribution of 
$u_{rms}/U_{j}$ along the lipline calculated by the current 
work and by the numerical reference present similar behavior. 
Nonetheless, the distributions are overestimated when compared 
to the experimental data.

%
%
%


\subsection{Dynamic Mode Decomposition Results}

The streaming version\cite{hemati2014dynamic} of the 
{\it total-least-squares} DMD algorithm\cite{hemati2015biasing} 
on volumetric data extracted during the large eddy simulations 
described in Sect.\ \ref{sec:LES} is computed in the present work.
Considering that the DMD calculation is performed in serial mode, the computer memory 
is the limiting factor to compute the DMD modes. The DMD 
calculation is run on a single processor with 128 GB of RAM\@.
According to Hemati {\it et al.}\cite{hemati2014dynamic}, the 
computational cost of the algorithm to calculate the DMD 
eigen-elements is $\mathcal{O}(nr^2)$, where $n$ and $r$ are the 
snapshot dimension and the maximum rank of the DMD operator, 
respectively. For the latter parameter, the streaming version of 
the DMD algorithm includes a compression step allowing to set it 
arbitrarily. Then, the choice of these parameters is a compromise 
between spatial and spectral resolution. The jet entrance, the 
potential core and the near field of the jet are included in the 
computational domain in order to prioritize the spatial aspects 
of the flow. 

Therefore, the results should include the aerodynamic 
structures as well as the generated acoustic waves. However, the 
original snapshots have been under-sampled in spatial resolution in
order to handle manageable snapshots. The dimensions of the snapshots
are specified in Tab.\ \ref{tab:simu}, counting approximatively 35 
million grid points. Finally, considering 256 snapshots without 
subtracting the mean, $r$ has been set equal to 50, which was the 
higher affordable number of retained modes in relation to the 
available computer memory.
In the present case, three variables were extracted from the LES calculations.
Hence, three different DMD reconstruction procedures were performed,
using snapshots of the velocity magnitude, the vorticity, based on the 
Q criterion, and the divergence of the velocity. In the following subsections, 
results are discussed regarding their spectral content 
(Sect.\ \ref{sec:spectral}) and spatial shape (Sect.\ \ref{sec:dmd-mode}).

\subsubsection{Spectral Analysis}\label{sec:spectral}

\begin{figure}[htb!]
  \centering
  \subfigure[Eigenvalues of $\mathcal{A}$, $\mu$]
     {\includegraphics[trim=1.0mm 2.0mm 2.0mm 1.0mm, clip,width=0.32\textwidth]
	 {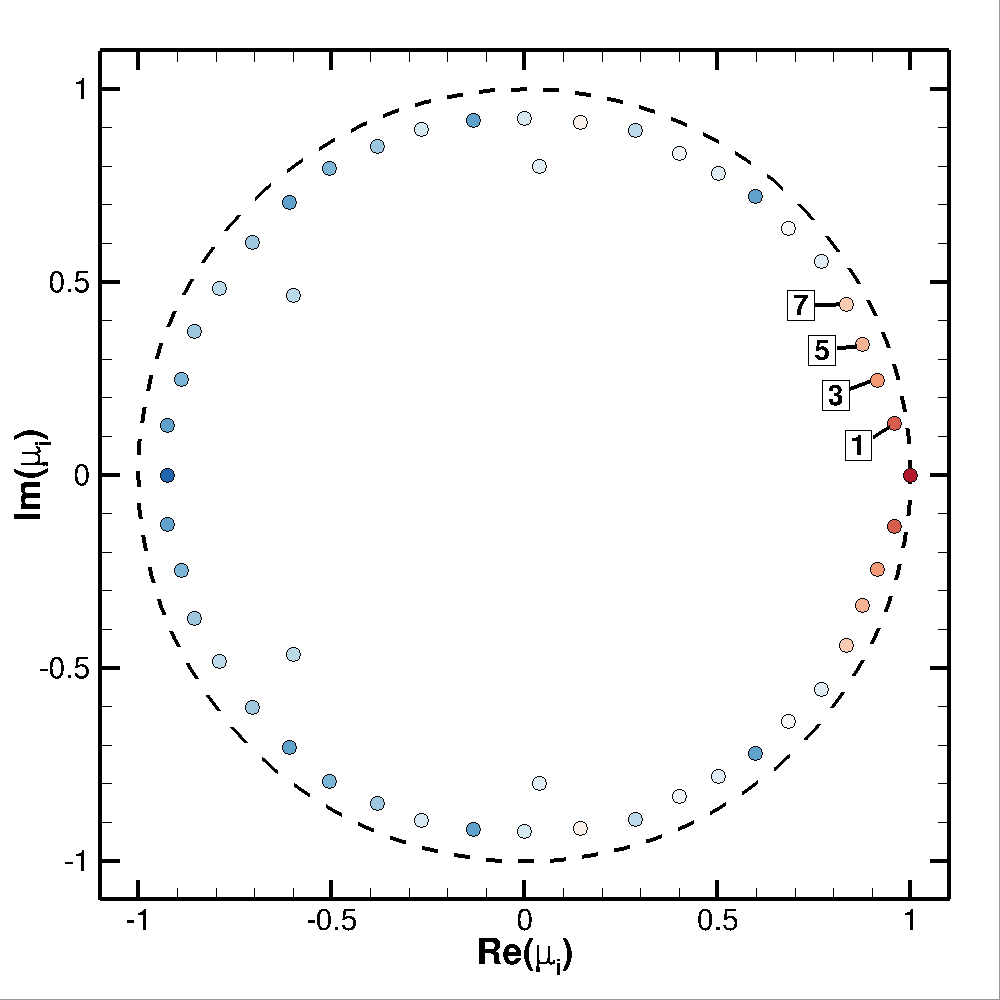}}
  \subfigure[Eigenvalues of the DMD modes, $\lambda$]
     {\includegraphics[trim=1.0mm 2.0mm 2.0mm 1.0mm, clip,width=0.32\textwidth]
	 {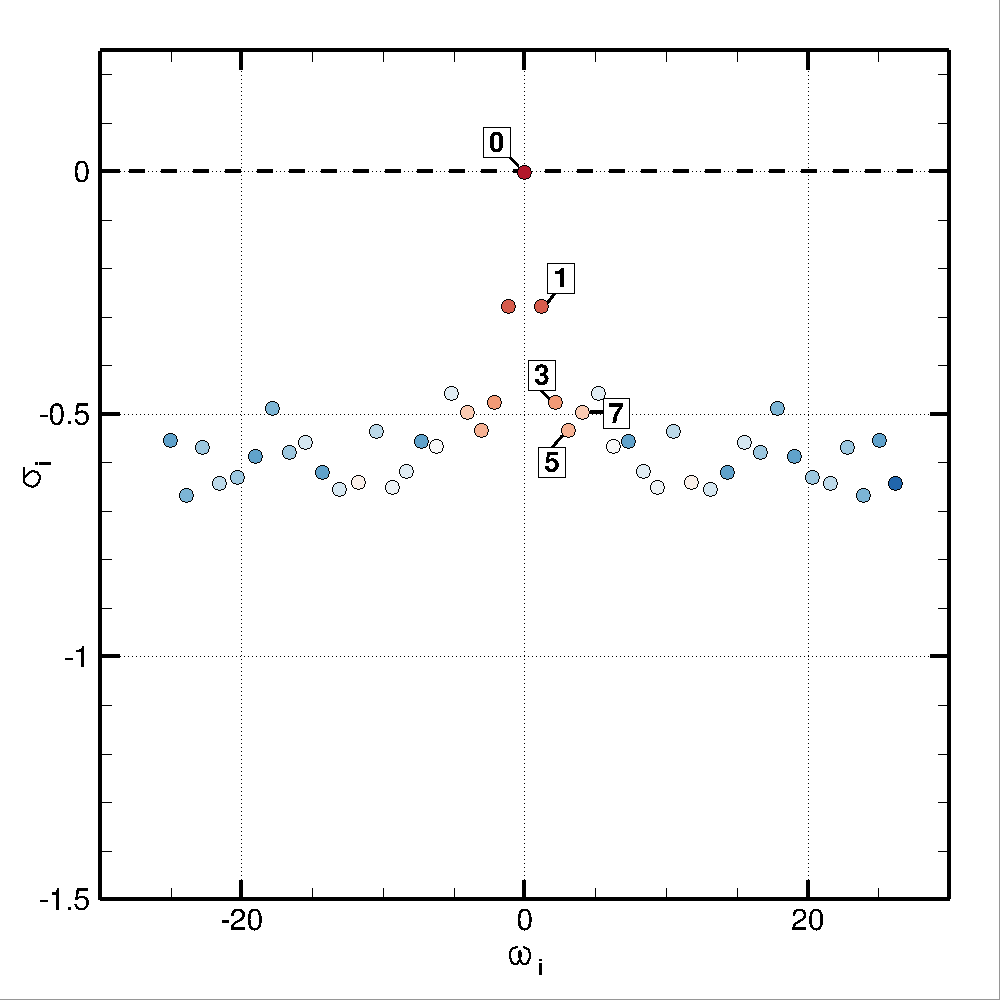}}
  \subfigure[Initial amplitude of the DMD modes, $\lVert \theta_i(1) \rVert$]
     {\includegraphics[trim=1.0mm 2.0mm 2.0mm 1.0mm, clip,width=0.32\textwidth]
	 {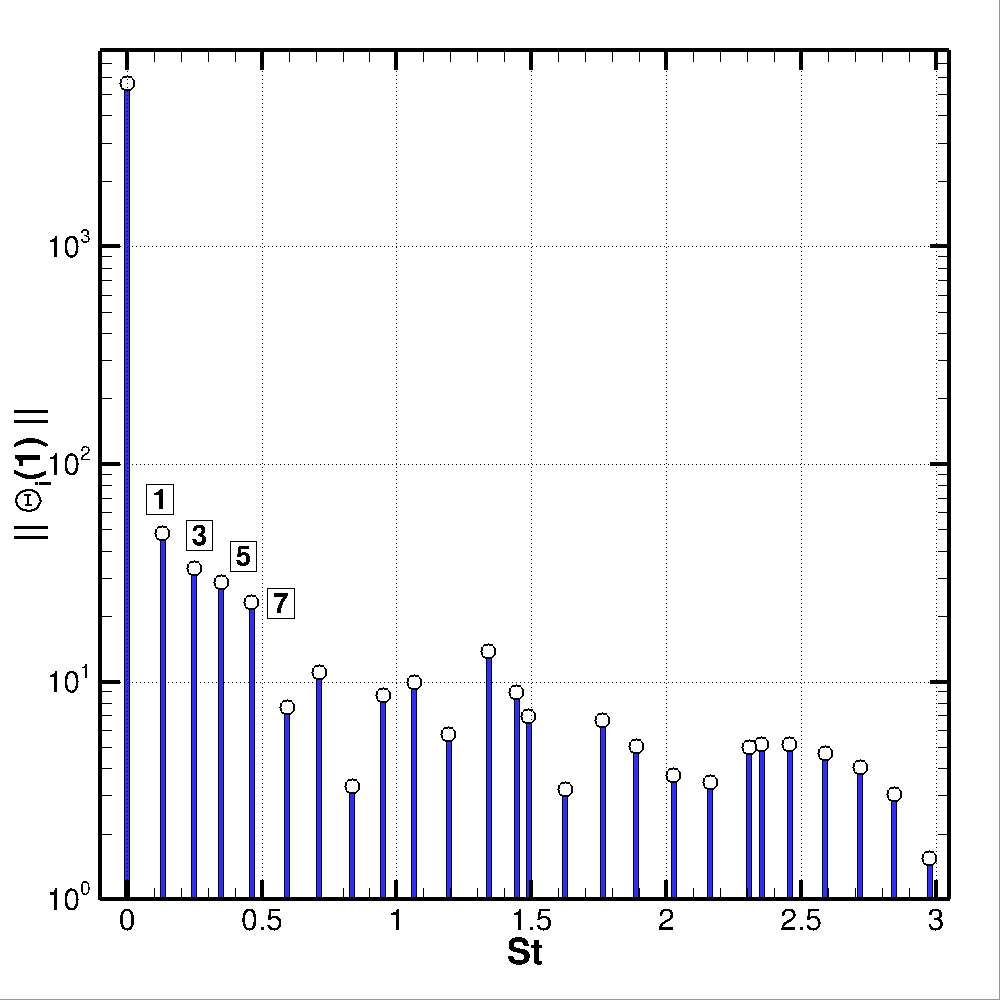}}
	\caption{Spectra from DMD computation using snapshots of velocity magnitude. In ($a$) and ($b$), 
	the symbols are colored by the mode amplitude, $\lVert \theta_{i}(1) \rVert$.}
	\label{fig:dmd-umag-spec}
\end{figure}

Figure \ref{fig:dmd-umag-spec} displays three different ways of 
representing the DMD spectrum obtained after the DMD computation 
using snapshots of the velocity magnitude. Figure 
\ref{fig:dmd-umag-spec}($a$) presents the 50 eigenvalues of 
$\mathcal{A}$ DMD linear operator. The symbols are colored by the 
initial amplitude of the DMD modes, $\lVert \theta_i(1) \rVert$, which 
are defined in Eq.\ \eqref{eq:theta1}. The choice of this parameter 
to differentiate the dynamic modes comes from Eqs.\ \eqref{eq:dmd0} and 
(\ref{eq:dmd3}). The initial amplitude of the DMD modes has also 
been taken into account by Sayadi {\it et al.} \cite{sayadi2013dynamic}.
All dynamic modes which are located inside the unit circle are stable. 
The only one DMD mode located on the unit circle is a steady mode which, 
in general, retrieves the mean characteristics of the flow 
\cite{chen2012variants}. The stable dynamic modes are unsteady and have 
a complex conjugate, symmetric with respect to the $\mbox{Im}(\mu_i)=0$ axis. 
In Fig.\ \ref{fig:dmd-umag-spec} ($b$), the growth rate of each mode, 
$\sigma_i$, is plotted versus the frequency, $\omega_i$. A mode is stable 
if $\sigma_i$ is negative, which is in agreement with the discussion considering 
Fig.\ \ref{fig:dmd-umag-spec}($a$). Finally, Fig.\ \ref{fig:dmd-umag-spec}
($c$) presents the most amplified DMD mode as a function of the Strouhal 
number. Four dynamic modes displaying a high amplitude have been selected. 
In Figs.\ \ref{fig:dmd-umag-spec}($b$) and ($c$), it appears that the 
stability of the mode is not linked with its initial amplitude. The DMD 
Mode 5 is more stable than the DMD Mode 7 ($\sigma_5 < \sigma_7$). However, 
$\lVert \theta_5(1)\rVert$ is larger than $\lVert \theta_7(1)\rVert$. Therefore, 
one can state that the dynamic mode 5 is initially more amplified than the 
dynamic mode 7, but it decays more quickly as the simulation advances in time.

\begin{figure}[htb!]
  \centering
  \subfigure[Eigenvalues of $\mathcal{A}$, $\mu$]
     {\includegraphics[trim=1.0mm 2.0mm 2.0mm 1.0mm, clip,width=0.32\textwidth]{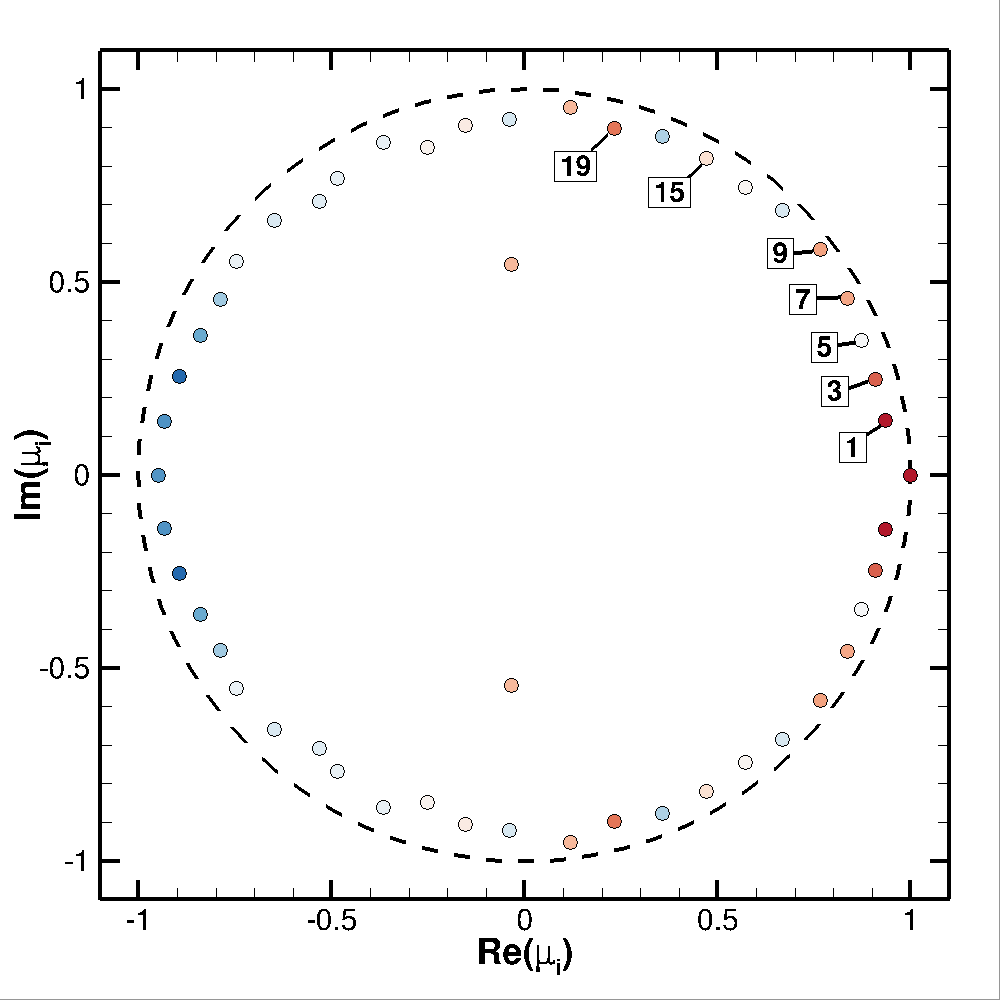}}
  \subfigure[Eigenvalues of the DMD modes, $\lambda$]
     {\includegraphics[trim=1.0mm 2.0mm 2.0mm 1.0mm, clip,width=0.32\textwidth]{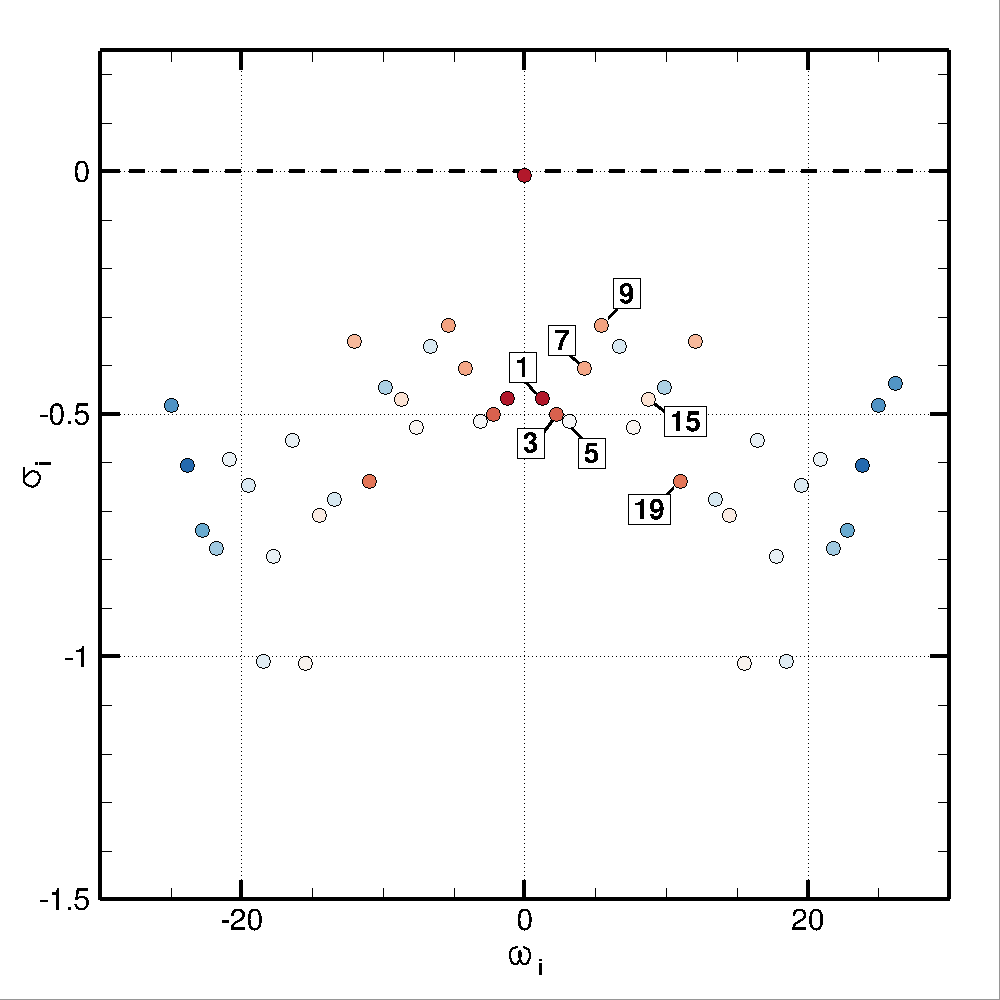}}
  \subfigure[Initial amplitude of the DMD modes, $\lVert \theta_i(1) \rVert$]
     {\includegraphics[trim=1.0mm 2.0mm 2.0mm 1.0mm, clip,width=0.32\textwidth]{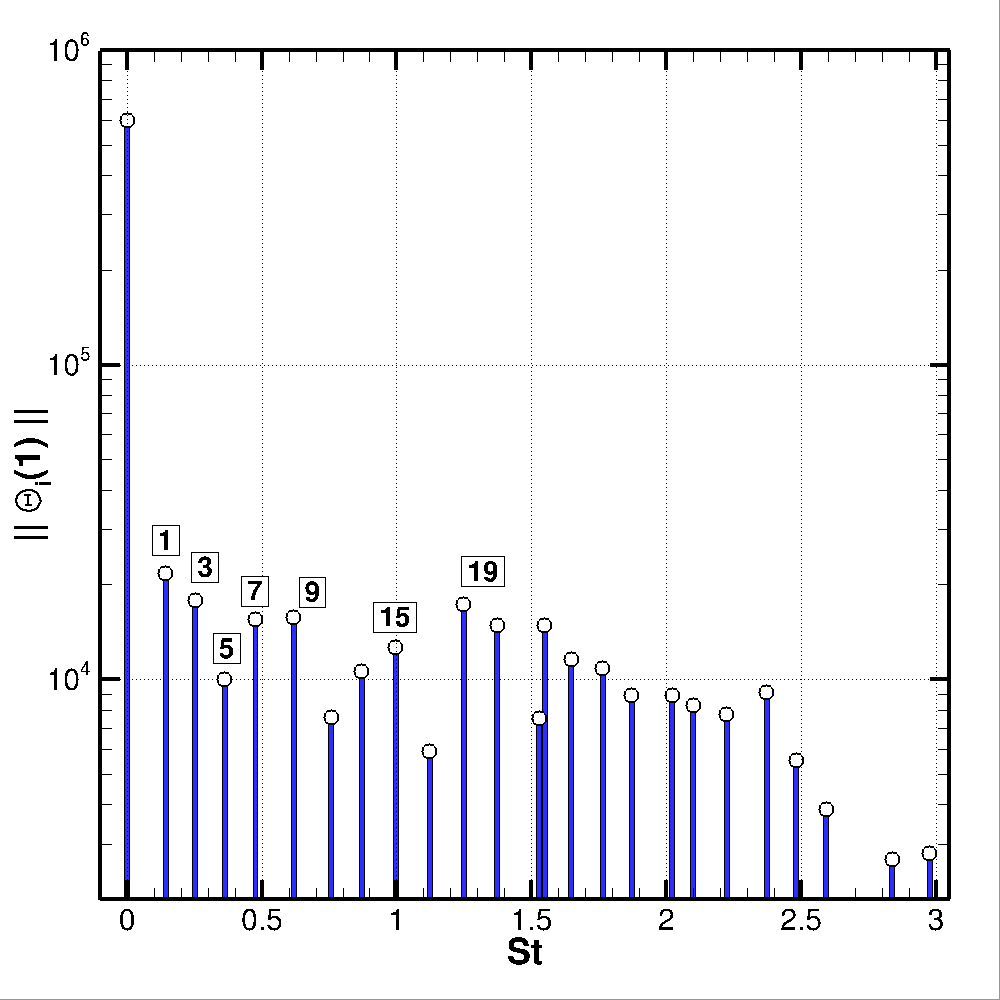}}
	\caption{Spectra from DMD computation using snapshots of vorticity. 
	In ($a$) and ($b$), the symbols are colored by the mode amplitude, $\lVert \theta_i(1) \rVert$.}
	\label{fig:dmd-vort-spec}
\end{figure}

\begin{figure}[htb!]
  \centering
  \subfigure[Eigenvalues of $\mathcal{A}$, $\mu$]
     {\includegraphics[trim=1.0mm 2.0mm 2.0mm 1.0mm, clip,width=0.32\textwidth]{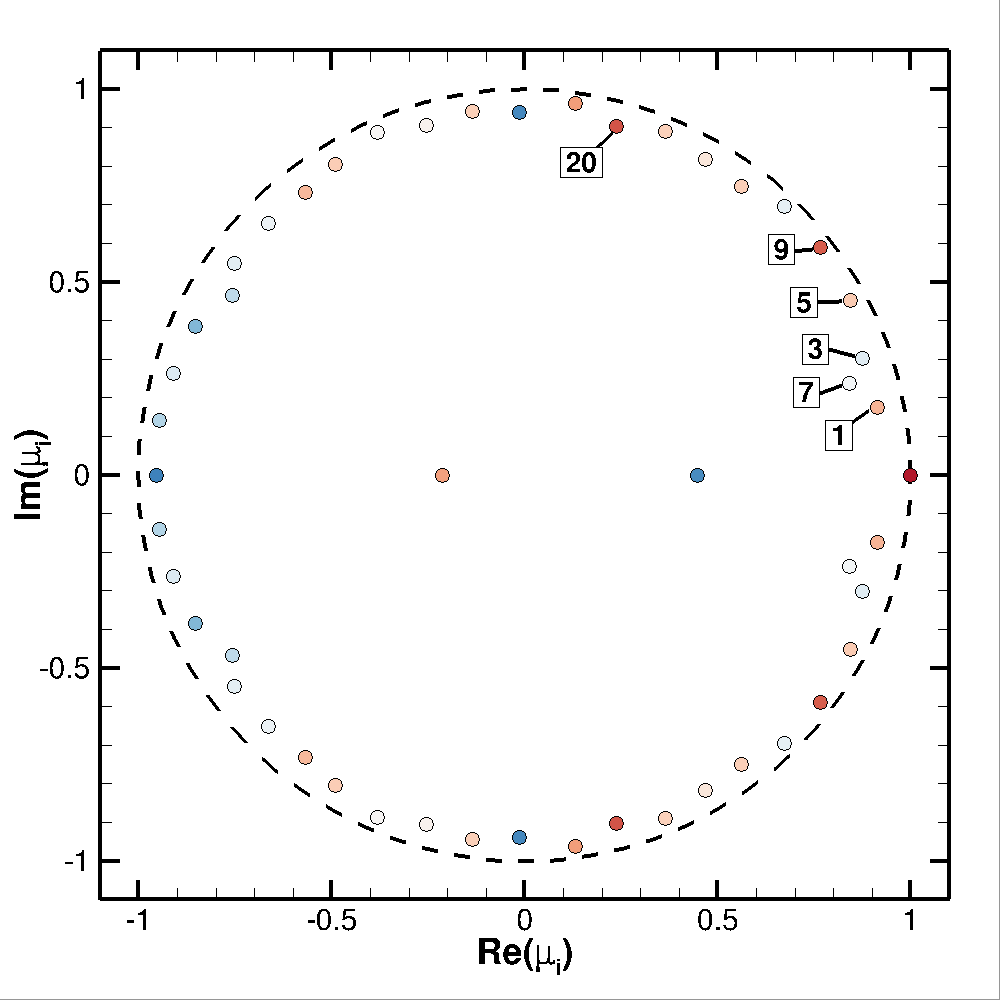}}
  \subfigure[Eigenvalues of the DMD modes, $\lambda$]
     {\includegraphics[trim=1.0mm 2.0mm 2.0mm 1.0mm, clip,width=0.32\textwidth]{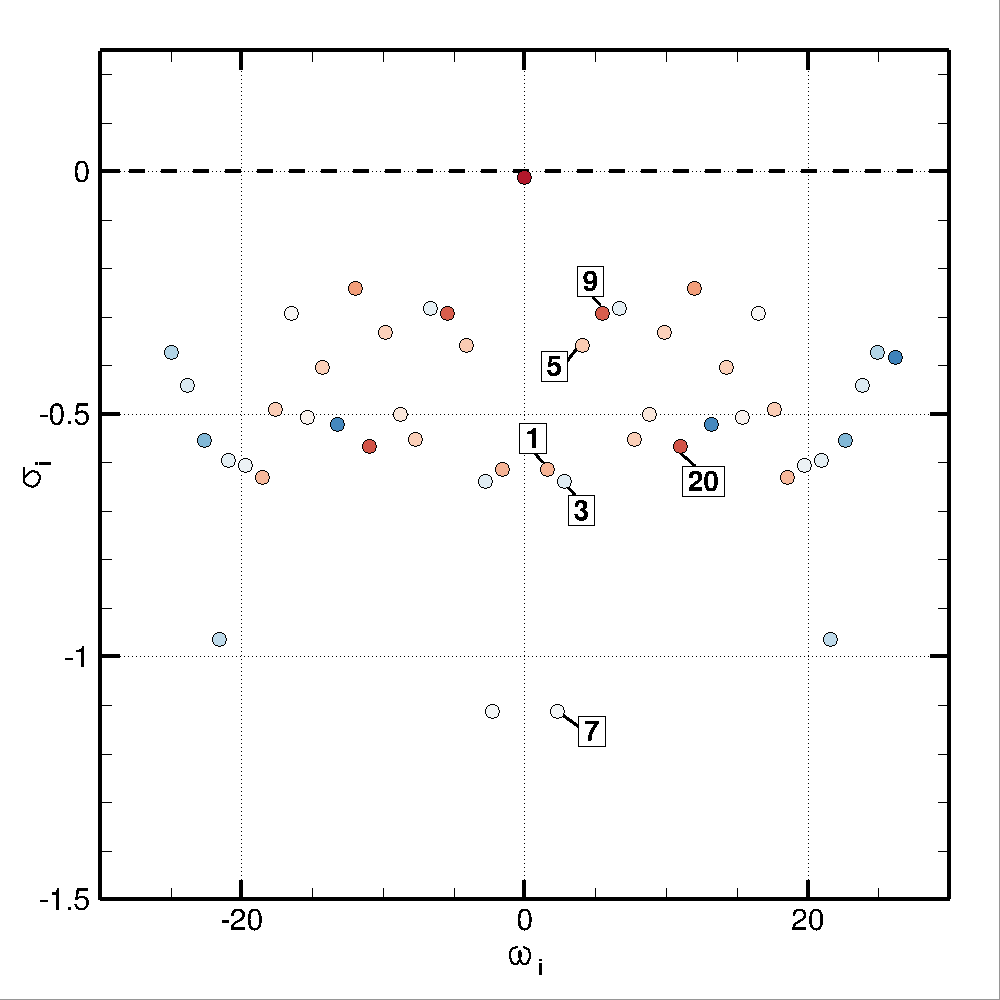}}
  \subfigure[Initial amplitude of the DMD modes, $\lVert \theta_i(1) \rVert$]
     {\includegraphics[trim=1.0mm 2.0mm 2.0mm 1.0mm, clip,width=0.32\textwidth]{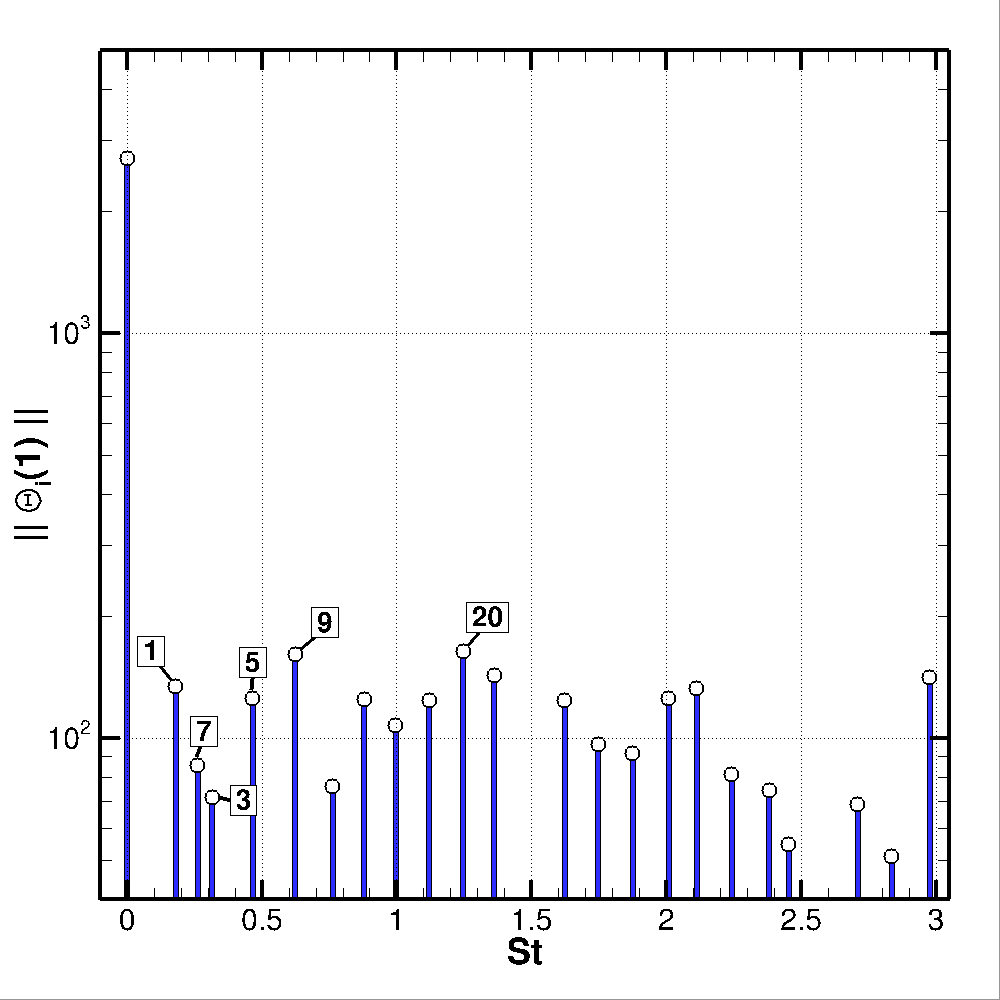}}
	\caption{Spectra from DMD computation using snapshots of divergence of velocity. 
	In ($a$) and ($b$), the symbols are colored by the mode amplitude, $\lVert \theta_i(1) \rVert$.}
	\label{fig:dmd-divv-spec}
\end{figure}

Figures \ref{fig:dmd-vort-spec} and \ref{fig:dmd-divv-spec} show two different 
sets of spectra obtained from the DMD computations using snapshots of the 
vorticity, based on the Q-criterion, and the divergence of the velocity, respectively.
Once again, all dynamic modes are stable, but the one representing the mean flow 
is neutrally stable. More DMD modes have been highlighted by a number in order 
to identify them in each spectrum.

One can observe in Figs.\ \ref{fig:dmd-umag-spec}($c$), \ref{fig:dmd-vort-spec}($c$)
and \ref{fig:dmd-divv-spec}($c$) that every spectra contain a dynamic mode at 
$St \approx 0.25$ and at $St \approx 0.48$. The clustering around specific frequencies 
for different DMD analyses denotes important dynamic activity at these frequencies.
The DMD modes associated to each frequency are shown in Tab.\ \ref{tab:freqmode}.
These characteristic frequencies coincide with the experimental far field pressure peaks 
observed by Bridges {\it et al.}\cite{bridges2008turbulence}. In the next subsection, 
the spatial shapes of these dynamic modes, given in Tab.\ \ref{tab:freqmode}, are discussed in 
more detail.
\begin{table}[htbp]
\begin{center}
	\caption{Characteristic frequencies and associated DMD modes}
\label{tab:freqmode}
\begin{tabular}{cccc}
\hline\hline
St & Velocity magnitude &  Vorticity & Divergence of velocity \\
\hline
0.25 & 3 & 3 & 7 \\
0.48 & 7 & 7 & 5 \\
\hline\hline
\end{tabular}
\end{center}
\end{table}
%

\subsubsection{Spatial Modes Analysis}\label{sec:dmd-mode}

\begin{figure}[htb!]
	\centering
    {\includegraphics[trim=1.0mm 2.0mm 2.0mm 1.0mm, clip,width=0.75\textwidth]
	{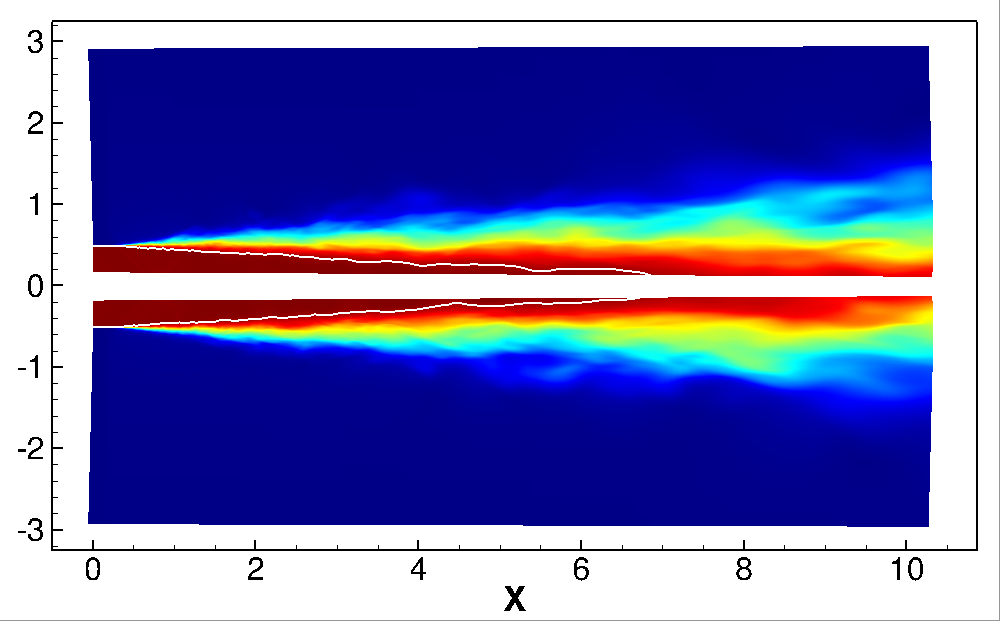}}
	\caption{Slice of the three dimensional steady DMD mode for the velocity 
	magnitude. The white line defines the potential core limits. Contours are 
	the same as in Fig.	\ref{fig:lat-u-av}.}
	\label{fig:dmd-umean}
\end{figure}

The averaged axial velocity component of the steady DMD mode is 
shown in Fig. \ref{fig:dmd-umean}, using the same color coding for the contours as the
LES mean flow illustrated in Fig. \ref{fig:lat-u-av}. One can notice 
a white gap around the centerline of the flow. The gap is created 
because the radial coordinate of the snapshot grid starts at the 20th
point in the radial direction, in order to reduce the computational 
cost of the DMD computation. The DMD mode has been reconstructed 
by multiplying the mode shape by its initial amplitude $\lVert \theta_0(1)
\rVert$. A fairly good agreement between the DMD calculation and the
large eddy simulation is found regarding the potential core 
length as well as the contour levels, even considering that the sample rate and the 
number of snapshots are quite different in the DMD calculation when compared to the LES 
statistics calculation.

\begin{figure}[htb!]
  \centering
  \subfigure[DMD mode 3 -- Velocity magnitude]
     {\includegraphics[trim=1.0mm 2.0mm 2.0mm 1.0mm, clip,width=0.45\textwidth]
	 {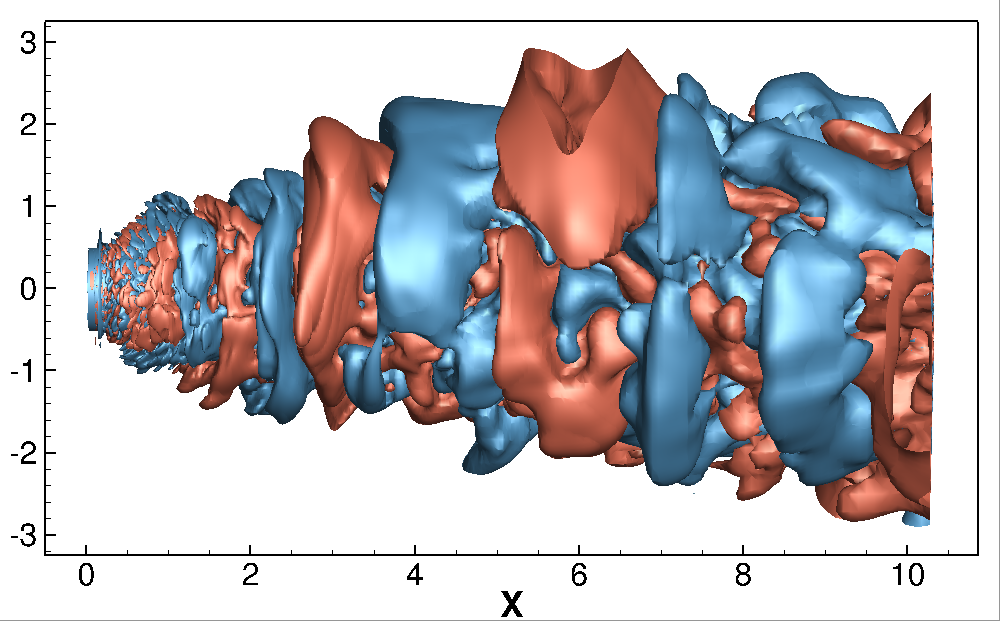}}
  \subfigure[DMD mode 3 -- Velocity magnitude]
     {\includegraphics[trim=1.0mm 2.0mm 2.0mm 1.0mm, clip,width=0.45\textwidth]
	 {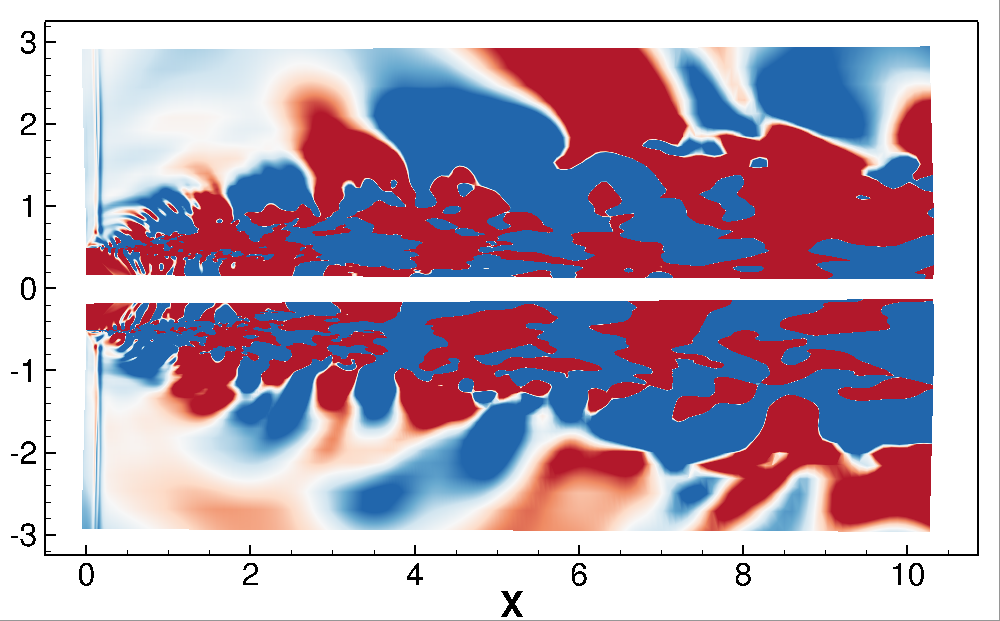}}\\
  \subfigure[DMD mode 3 -- Vorticity (Q criterion)]
     {\includegraphics[trim=1.0mm 2.0mm 2.0mm 1.0mm, clip,width=0.45\textwidth]
	 {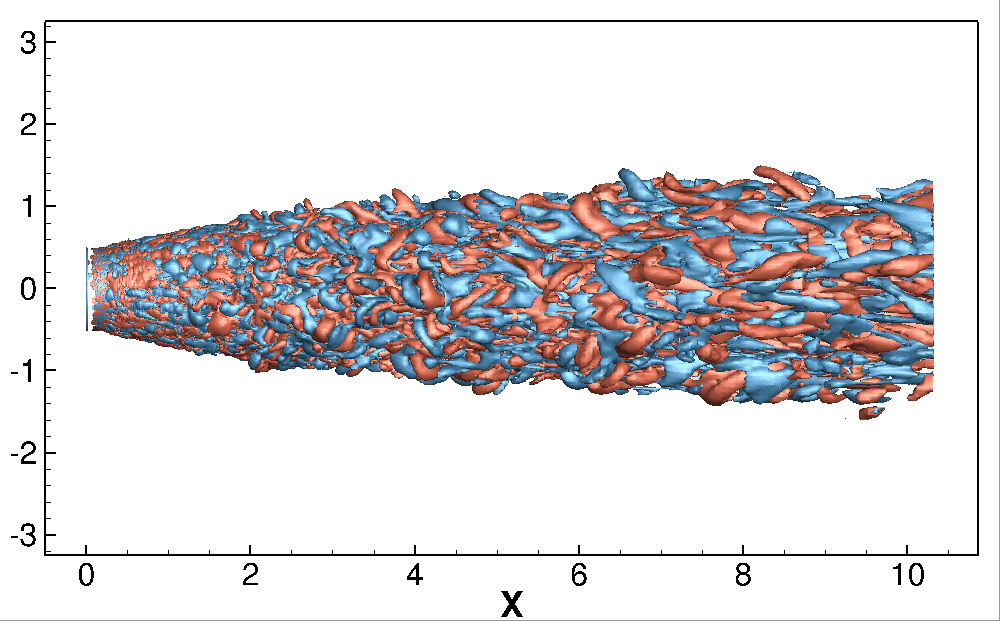}}
  \subfigure[DMD mode 3 -- Vorticity (Q criterion)]
     {\includegraphics[trim=1.0mm 2.0mm 2.0mm 1.0mm, clip,width=0.45\textwidth]
	 {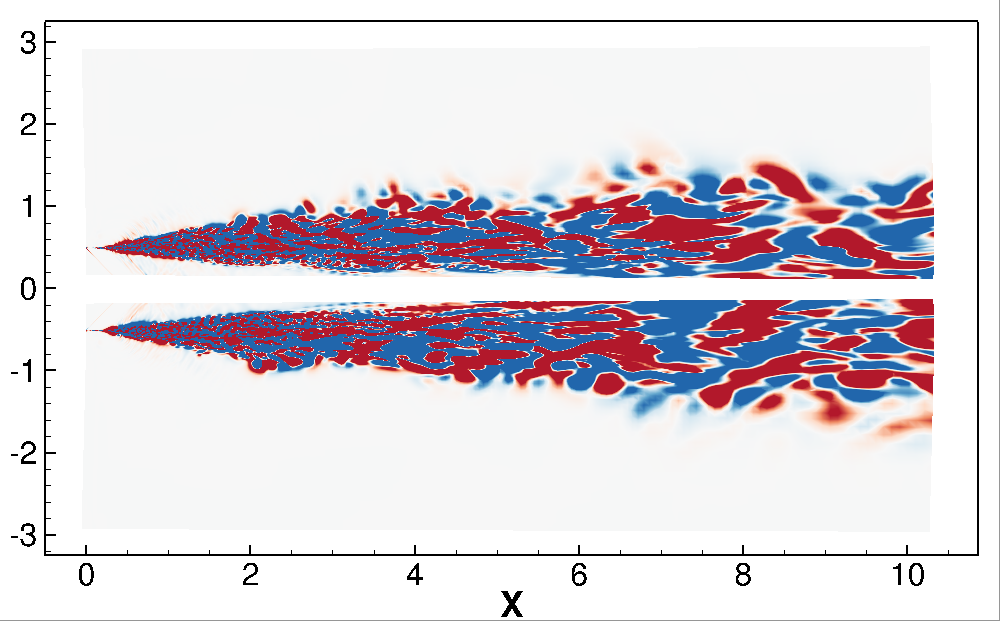}}\\
  \subfigure[DMD mode 7 -- Divergence of velocity]
     {\includegraphics[trim=1.0mm 2.0mm 2.0mm 1.0mm, clip,width=0.45\textwidth]
	 {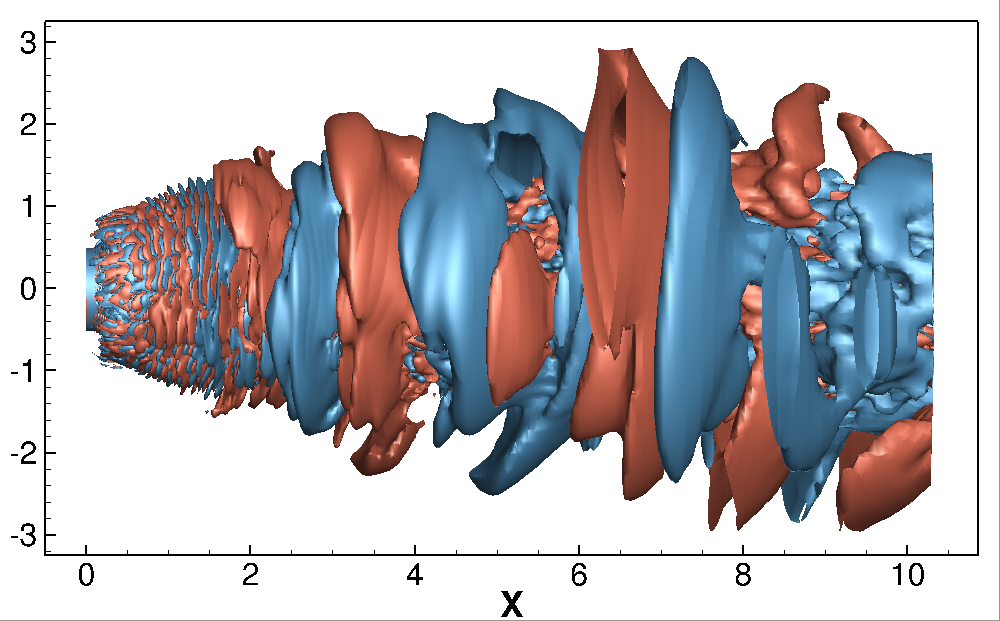}}
  \subfigure[DMD mode 7 -- Divergence of velocity]
     {\includegraphics[trim=1.0mm 2.0mm 2.0mm 1.0mm, clip,width=0.45\textwidth]
	 {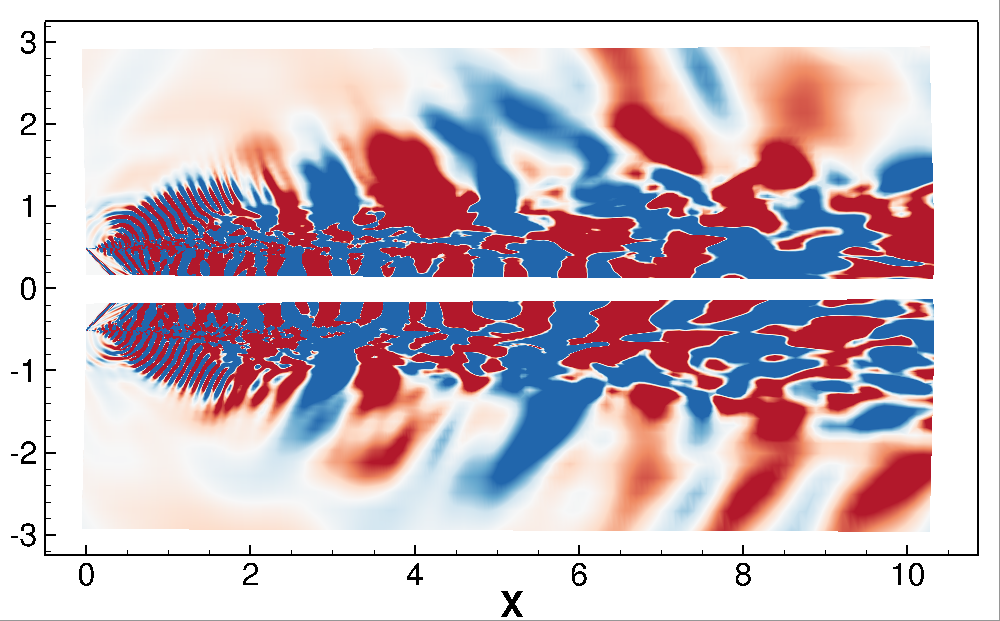}}\\
	\caption{Visualization of the DMD modes found at $St \approx 0.25$\@. 
	($a$) and ($b$) display the real part of Mode 3 extracted from the DMD analysis 
	using snapshots of velocity magnitude, ($c$) and ($d$) display the real part 
	of Mode 3 extracted from the DMD analysis using snapshots of vorticity (Q criterion), and ($e$) 
	and ($f$) display the real part of Mode 7 extracted from the DMD analysis using 
	snapshots of divergence	of velocity. The left and right columns show 3-D 
	isosurfaces and 2-D cut-plane visualizations of the modes, respectively (positive in 
	red and negative in blue).}
	\label{fig:dmd-mode-1}
\end{figure}

\begin{figure}[htb!]
  \centering
  \subfigure[DMD mode 7 -- Velocity magnitude]
     {\includegraphics[trim=1.0mm 2.0mm 2.0mm 1.0mm, clip,width=0.45\textwidth]
	 {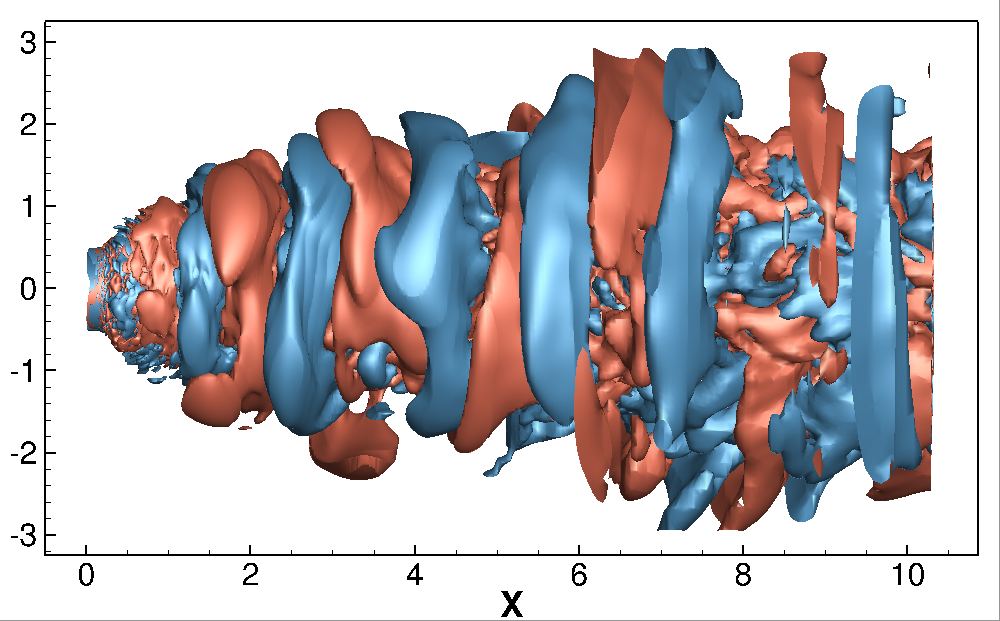}}
  \subfigure[DMD mode 7 -- Velocity magnitude]
     {\includegraphics[trim=1.0mm 2.0mm 2.0mm 1.0mm, clip,width=0.45\textwidth]
	 {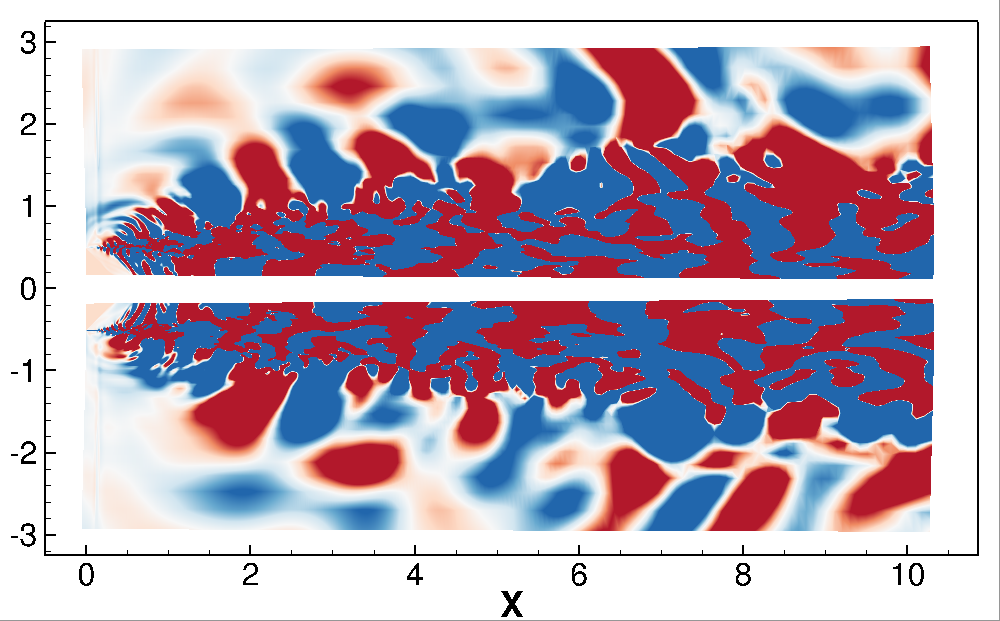}}\\
  \subfigure[DMD mode 7 -- Vorticity (Q criterion)]
     {\includegraphics[trim=1.0mm 2.0mm 2.0mm 1.0mm, clip,width=0.45\textwidth]
	 {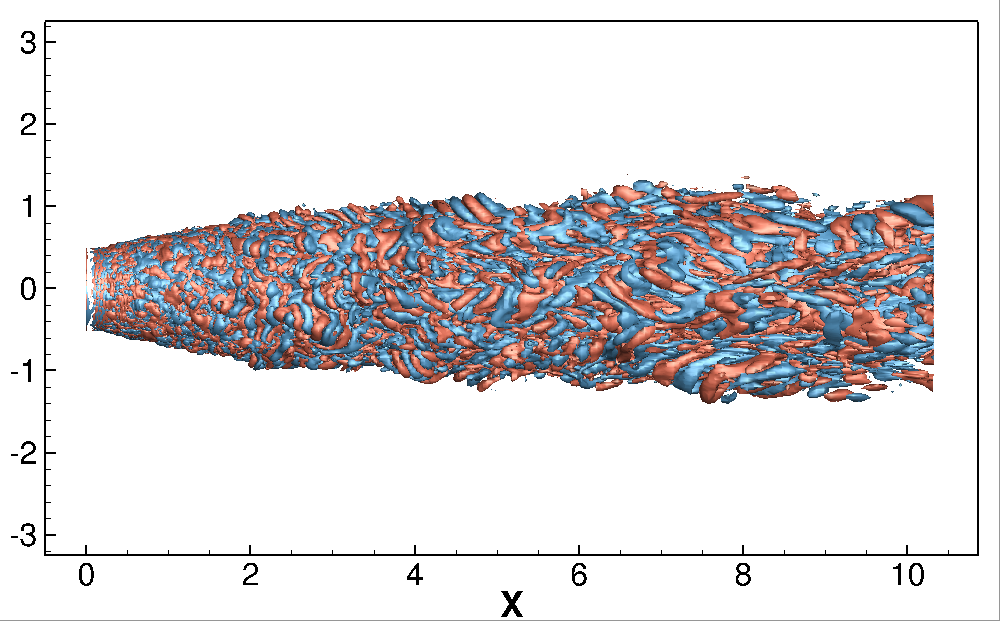}}
  \subfigure[DMD mode 7 -- Vorticity (Q criterion)]
     {\includegraphics[trim=1.0mm 2.0mm 2.0mm 1.0mm, clip,width=0.45\textwidth]
	 {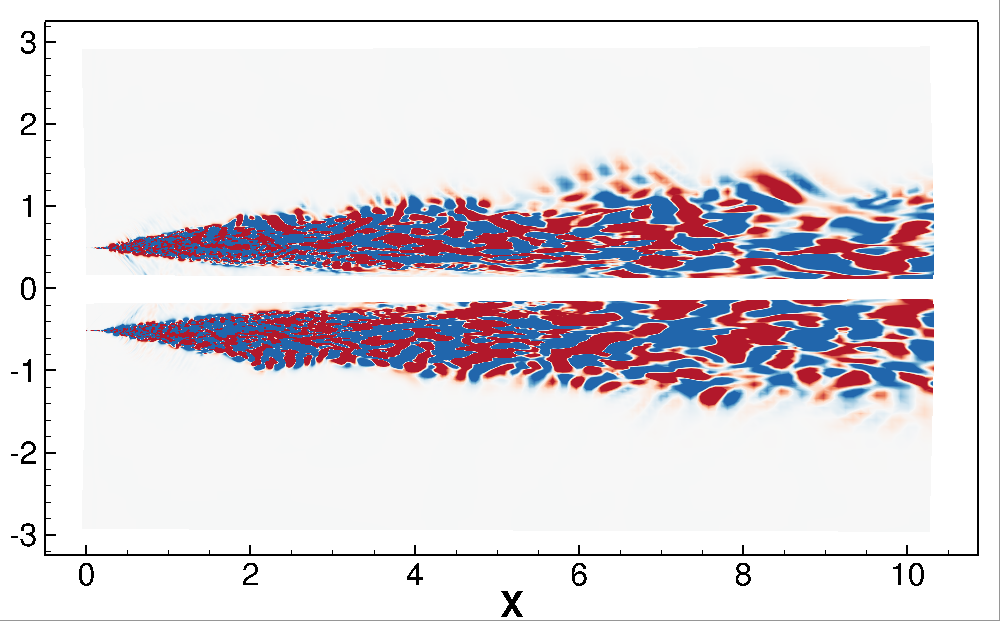}}\\
  \subfigure[DMD mode 5 -- Divergence of velocity]
     {\includegraphics[trim=1.0mm 2.0mm 2.0mm 1.0mm, clip,width=0.45\textwidth]
	 {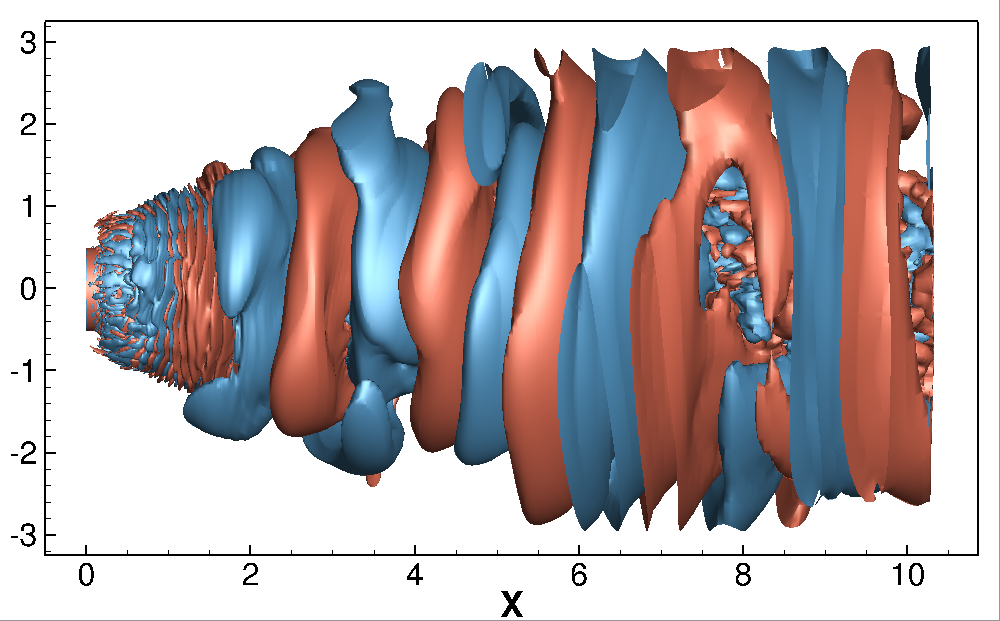}}
  \subfigure[DMD mode 5 -- Divergence of velocity]
     {\includegraphics[trim=1.0mm 2.0mm 2.0mm 1.0mm, clip,width=0.45\textwidth]
	 {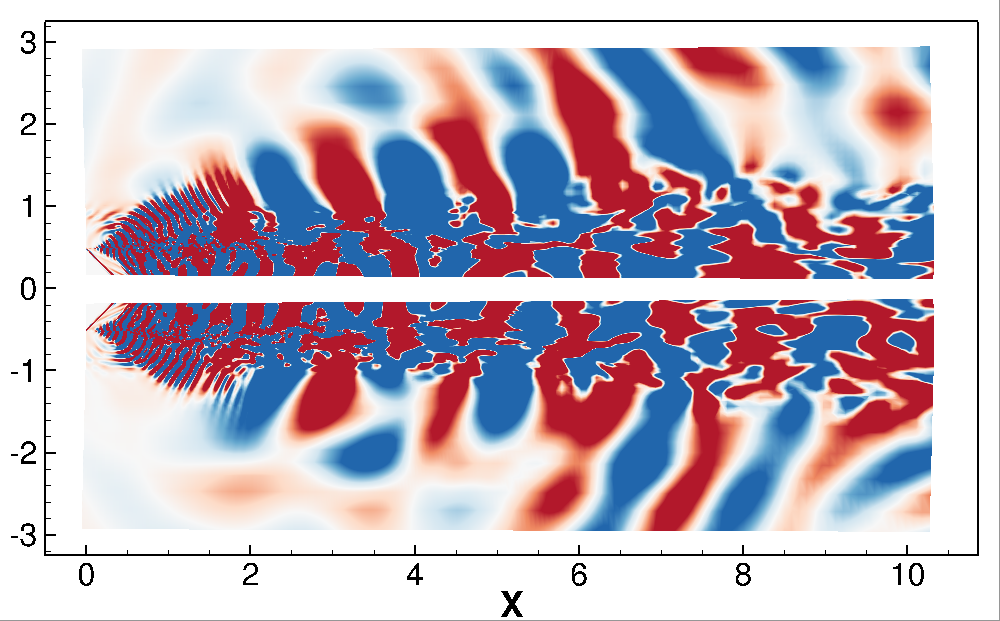}}\\
	\caption{Visualization of the DMD modes found at $St \approx 0.48$\@. 
	($a$) and ($b$) display the real part of Mode 7 extracted from the DMD analysis 
	using snapshots of velocity magnitude, ($c$) and ($d$) display the real part of 
	Mode 7 extracted from the DMD analysis using snapshots of vorticity (Q criterion), and ($e$) 
	and ($f$) display the real part of Mode 5 extracted from the DMD analysis using 
	snapshots of divergence	of velocity. The left and right columns show 3-D isosurfaces 
	and 2-D cut-plane visualizations of the modes, respectively (positive in red and 
	negative in blue).}
	\label{fig:dmd-mode-2}
\end{figure}

As mentioned in the previous subsection, the experimental far field pressure spectrum 
of Bridges {\it et al.}\cite{bridges2008turbulence} displays two peaks at 
$St \approx 0.25$ and $St \approx 0.48$\@. Modes at the same frequencies are observable in the 
three DMD analyses performed in the present study. Figure \ref{fig:dmd-mode-1} displays the DMD 
modes associated to the first frequency, $St \approx 0.25$, while Fig.\ 
\ref{fig:dmd-mode-2} shows the DMD modes associated to $St \approx 0.48$. In 
both figures, isosurfaces and 2-D cut planes of velocity magnitude, 
vorticity (Q criterion) and divergence of velocity are presented. Considering the high Reynolds 
number of the present work, and the rapid transition from laminar flow at the 
jet inlet to a turbulent jet mixing layer, it is possible to observe 
coherent behavior in the jet dynamics. Moreover, the three variables, for which  
the DMD computations were performed, bring different information about the flow dynamics. 
While the vorticity modes seem to enlighten the mixing layer dynamics, the velocity magnitude 
as well as the divergence of velocity seem to highlight the aeracoustic dynamics.

Figures \ref{fig:dmd-mode-1} and \ref{fig:dmd-mode-2} indicate that, until $x \approx 1$, 
small coherent vortical structures are growing in the jet mixing layer, generating small acoustic 
waves. Further downstream, the flow has already transitioned and large acoustic waves are generated 
and are propagated in the downstream direction. As expected, the wavelength of the large acoustic 
waves depends on the DMD mode frequency. One can see, when comparing, for instance, 
Fig.\ \ref{fig:dmd-mode-1}($a$) with Fig.\ \ref{fig:dmd-mode-2}($a$), or Fig.\ \ref{fig:dmd-mode-1}($e$) 
with Fig.\ \ref{fig:dmd-mode-2}($e$), that the wavelength of the coherent structures is divided by two
when the frequency is doubled. Moreover, it is easy to verify, for instance, in 
Fig.\ \ref{fig:dmd-mode-2}($f$), the relation between the wavelength of the large acoustic 
waves and the actual frequency of the DMD mode, $\omega_i$.

Another interesting aspect is the presence of small vortices in the inner mixing layer,
at the interface with the potential core. These structures are visible in 
Figs.\ \ref{fig:dmd-mode-1}($d$) and \ref{fig:dmd-mode-2}($d$). Unfortunately, due to the 
absence of grid points along the centerline itself in the grid used to extract the data for 
in the present DMD calculations, the influence of these small vortices at the end of the potential 
core is not accessible in the present case. Future work should consider a snapshot grid covering 
all the inner part of the jet.
%
%
Finally, one can see in Fig.\ \ref{fig:dmd-mode-2}($c$) that the vortex filaments in the mixing 
layer seem to suffer a three-dimensional helicoidal distortion around the jet mixing layer.
The work of Violato and Scarano \cite{violato2013three}, who performed experiments for 
a low Reynolds free water jet, using time-resolved tomographic particle image velocimetry 
(TR-TOMO PIV), can certainly help in the understanding of this type of fundamental aspect in 
the current jet dynamics.


  \section{Concluding Remarks}

The present work is concerned with the study of the aerodynamics 
of a perfectly expanded supersonic jet flow. It is expected that the 
flow data and the reduced order model here generated could be used in the 
future for performing aeroacoustic studies of jet flows. 
An implicit large eddy simulation (LES) formulation for 
compressible flows, based on the System I set of equations, is 
used. A streaming version of the total-least-squares DMD algorithm
is chosen to run concurrently with the LES simulation and 
provide an additional form of studying the more relevant aspects of the jet dynamics.

LES of a high Reynolds perfectly expanded supersonic jet flow configuration is performed
on a computational mesh with 85 million grid points. Statistical data are extracted
from the simulations and present good agreement with the numerical and experimental 
reference work, at least near the jet inlet region where the mesh is well refined.
However, this is not the case when the jet moves away from the domain entrance. 
As a result, the potential core length calculated by the present LES is underestimated.
Such behavior could be expected since the low order numerical scheme of the numerical 
solver presently used would probably require quite extensive mesh refinements. 
The work also presents three DMD analyses, which have been performed by extracting large 
three-dimensional snapshots from the LES results. These DMD computations concerned
the velocity magnitude, the vorticity, based on the Q criterion, and the divergence 
of the velocity. Two frequencies are identified for which all DMD calculations identify
a dynamic mode with relevant flow structures. These frequencies agree with those of relevant 
dynamics identified in previous experimental work available in the literature. 
The analysis of all the dynamic modes brought new insights on the jet dynamics
regarding the vortical structures and the acoustic wave patterns.

At the time of this writing, the LES solver is being adapted in order to include parallel I/O
features. This capability will open new opportunities in term of additional grid resolution
that would allow a reduction in the difference between the results calculated by the authors 
and other data, computational or experimental, available in the literature. 
Moreover, the DMD algorithm here implemented should also be parallelized in order to allow 
handling larger snapshots and, hence, the extraction of more information from the flow, 
especially at the centerline of the jet and further downstream of the jet entrance. Hopefully,
these modifications will allow sufficient mesh refinement, both for the LES calculations and 
for DMD analyses, that the present tool will be useful for studies of the jet aeroacoustics. 


\section*{Acknowledgments}

The authors gratefully acknowledge the partial support for this research provided by
Conselho Nacional de Desenvolvimento Cient\'{\i}fico e Tecnol\'{o}gico, CNPq, under 
the Research Grants No.\ 309985/2013-7, No.\ 400844/2014-1, No.\ 443839/2014-0 and 
No.\ 150450/2016-8\@. The authors are also indebted to the partial financial support 
received from Funda\c{c}\~{a}o de Amparo \`{a} Pesquisa do Estado de S\~{a}o Paulo, 
FAPESP, under the Research Grants No.\ 2013/07375-0 and No.\ 2013/21535-0. 


\bibliography{sources/references}
\bibliographystyle{aiaa}

\end{document}